\documentclass[english]{article} 
\usepackage[margin=1in]{geometry} 
\usepackage{babel} 
\usepackage{xcolor}

\usepackage{setspace} 
\usepackage{lipsum} 
\usepackage{graphicx} 

\usepackage{threeparttable}

\usepackage{amsmath} 
\usepackage{amssymb} 
\usepackage{amsfonts} 

\usepackage{caption} 
\usepackage{subcaption} 

\usepackage{booktabs} 

\usepackage{enumitem} 

\usepackage{float} 

\usepackage{hyperref} 
\hypersetup{
    colorlinks = true, 
    linkcolor = blue, 
    citecolor = green, 
    urlcolor = cyan, 
    bookmarksopen = true, 
    bookmarksnumbered = true, 
    pdftitle = {Study of Neutron Star Properties under the Two-Flavor Quark NJL Model}, 
    pdfauthor = {Chunran Zhu, Bolin Li} 
}

\usepackage[backend=bibtex, citestyle=numeric, bibstyle=numeric, sorting=none]{biblatex} 

\usepackage{threeparttable} 
\usepackage{pdflscape} 

\title{Study of Neutron Star Properties under the Two-Flavor Quark NJL Model}
\author{
  \begin{tabular}{c}
    Chun-Ran Zhu\thanks{Email: zhuchunran@outlook.com} \quad Bo-Lin Li\thanks{Corresponding author email: blli@usst.edu.cn} \\
    \addlinespace[1.0em]
    \normalsize  School of Physics, Faculty of Basic Sciences, University of Shanghai for Science and Technology \\
    \normalsize Shanghai 200093, China
  \end{tabular}
}
\date{\today}

\begin{document}

\sloppy 

\maketitle 
\thispagestyle{empty} 

\begin{abstract}
    The Equation of State (EOS) of matter within neutron stars is a central topic in nuclear physics and astrophysics.
    This study investigates hadron-quark hybrid stars by integrating the density-dependent DDME2 relativistic mean-field model for hadronic matter with a two-flavor Nambu-Jona-Lasinio (NJL) model for quark matter.
    A quintic polynomial interpolation is employed to construct a smooth ($C^2$ continuity) and thermodynamically consistent crossover between the phases.
    We systematically explore the parameter space to reconcile the tension between the high stiffness required by massive pulsars and the softness demanded by tidal deformability and radius constraints.
    Our analysis demonstrates that to simultaneously satisfy the mass measurement of PSR J0740+6620 and the compact radius constraints from NICER (e.g., PSR J0437-4715), the hadron-quark crossover must initiate in the vicinity of nuclear saturation density.
    This result suggests that the early percolation of quark degrees of freedom is a necessary feature to accommodate current multi-messenger observations.

    \vspace{1em} 

\end{abstract}

\clearpage 

\section{Introduction}
\label{sec:introduction}
    Neutron stars are extremely dense celestial objects formed from the gravitational collapse of massive stars. Their core densities can reach several times the nuclear saturation density, providing a unique natural laboratory for studying strongly interacting matter under extreme conditions \autocite{Baym_1975_NeutronStars, Baym_1979_PhysicsNeutronStars}. Consequently, precisely determining the Equation of State (EOS) of their internal matter is a key challenge at the intersection of nuclear physics and astrophysics, as it directly governs the macroscopic structure and properties of neutron stars \autocite{Baym_2018_FromHadronsToQuarks, Li_2022_PropertiesOfHybridStars, Li_2022_HybridStarsQuarkCore}.

    In recent years, advances in multi-messenger astronomy have provided unprecedented opportunities to constrain the EOS, while also revealing its inherent complexity. On one hand, precise observations of massive pulsars, particularly PSR J0740+6620 ($M \approx 2.08 M_{\odot}$) \autocite{Salmi_2024_PSRJ0740, Cromartie_2019}, require the EOS to be sufficiently "stiff" at high densities to support neutron stars exceeding two solar masses \autocite{Demorest_2010_TwoSolarMass, Fonseca_2016_NANOGrav, Kojo_2015_PhenomenologicalQCD}. On the other hand, the tidal deformability parameter ($\Lambda$) inferred from the gravitational wave signal of the binary neutron star merger GW170817 demands that the EOS for a $1.4 M_{\odot}$ neutron star be relatively "soft" at corresponding densities, with $\Lambda_{1.4} \lesssim 800$ \autocite{Abbott_2017_GW170817, Abbott_2017_MultiMessenger, Abbott_2018_GW170817EOS, Li_2022_HybridStarsQuarkCore, Abbott_2019_GW170817Props}. 
    
    Furthermore, the landscape of observational constraints has been significantly enriched by precise radius measurements from the Neutron Star Interior Composition Explorer (NICER). Observations of pulsars such as PSR J0030+0451 \autocite{Riley_2019_PSRJ0030} and PSR J0437-4715 \autocite{Bogdanov_2019_J0437} have provided simultaneous mass and radius estimates, generally favoring relatively compact radii in the $\sim 11-13$ km range for stars near $1.4 M_{\odot}$. This introduces a third crucial benchmark: a successful EOS must not only be stiff enough at high densities to support massive pulsars and soft enough at intermediate densities to satisfy tidal deformability limits, but also simultaneously predict radii consistent with these new, more compact NICER data. Reconciling this multi-faceted tension---high-density stiffness versus intermediate-density softness---within a unified physical framework is a central challenge in neutron star physics.

    Reconciling this multi-faceted tension---high-density stiffness, intermediate-density softness, and specific radius constraints---within a unified physical framework is a central problem in neutron star physics.

    A possible solution is to introduce a phase transition from hadronic matter to quark matter in the core of the neutron star, forming what is known as a "hybrid star" \autocite{Baym_2018_FromHadronsToQuarks, Li_2022_HybridStarsQuarkCore, Li_2022_PropertiesOfHybridStars}. The theoretical description of such a hybrid star requires distinct frameworks for the hadronic and quark phases. For the hadronic phase, various approaches exist, including phenomenological models like Skyrme or Relativistic Mean-Field (RMF) theory \autocite{Reinhard_1989_RMFReview}, as well as microscopic models based on Chiral Effective Field Theory (EFT). RMF models, in particular, have been highly successful in describing the properties of finite nuclei and nuclear matter saturation. For the high-density quark phase, the non-perturbative nature of Quantum Chromodynamics (QCD) necessitates the use of effective models. While simple approaches like the MIT Bag Model provide a basic description, they fail to capture key QCD dynamics. The Nambu-Jona-Lasinio (NJL) model, which we employ in this study, offers a more sophisticated framework \autocite{Nambu_1961, Buballa_2005_NJLReview}. As an effective theory, it can successfully describe spontaneous chiral symmetry breaking and its restoration at high density, which is a crucial non-perturbative feature expected in dense quark matter \autocite{Hatsuda_1994_QCDPhenomenology}. By combining a well-established RMF model for the hadronic phase with the NJL model for the quark phase, we construct a complete hybrid equation of state. We pay special attention to the effects of \textbf{quark vector interactions} within the model, as this interaction provides the necessary repulsion at high densities and is a key mechanism for stiffening the EOS to support massive neutron stars \autocite{Kojo_2015_PhenomenologicalQCD, Yuan_2022_InteractingQuarkMatter}. 
    
    Equally important is the method used to join the two phases. Instead of a sharp, first-order phase transition (e.g., a Maxwell construction), which often results in an EOS that is too soft to support massive neutron stars, we implement a smooth hadron-quark crossover. This is achieved through a thermodynamically consistent quintic polynomial interpolation that ensures $C^2$ continuity (continuous pressure, number density, and susceptibility) \autocite{Baym_2018_FromHadronsToQuarks, Li_2022_HybridStarsQuarkCore}. This crossover approach not only avoids unphysical jumps but also provides a more realistic representation of the transition, which is expected to be continuous at the high densities relevant to neutron stars.
    
    The goal of this research is to systematically explore the parameter space of this RMF-NJL hybrid model with a $C^2$ crossover. We specifically investigate the roles of the quark vector interaction ($G_V$) and the width of the crossover region ($BU$) to determine if this framework can construct an EOS that is not only theoretically self-consistent but also capable of simultaneously passing the stringent, multi-messenger tests from massive pulsars, gravitational waves, and NICER observations.

    In the subsequent sections, we present a detailed description of the NJL quark model and the DDME2 hadronic model, along with the interpolation method used to ensure $C^2$ continuity of the EOS. We then systematically analyze the impact of three key model parameters---the vector coupling constant $G_V$, the phase transition endpoint $BU$, and the scalar coupling factor $G_S\Lambda^2$---on the macroscopic properties of neutron stars. Through this analysis, we establish a fiducial benchmark set and discuss its consistency with astronomical observations. Finally, we summarize the findings of this paper.

\section{NJL Model and Quark Matter Equation of State}
\label{sec:njl-model}
    The Nambu-Jona-Lasinio (NJL) model is an effective quantum field theory used to describe the strong interactions between quarks, particularly well-suited for studying the phase transition from hadronic to quark matter in nuclear matter and dense stars \autocite{Buballa_2005_NJLReview, Nambu_1961}. It offers distinct advantages in describing phenomena such as spontaneous chiral symmetry breaking and restoration, the generation of dynamical quark masses, and the density-dependent nature of quark masses \autocite{Buballa_2005_NJLReview, Hatsuda_1994_QCDPhenomenology}. This section will detail the fundamental structure of the two-flavor NJL model, its parameter choices, and the calculation of the quark matter EOS at zero temperature and finite chemical potential, thereby establishing a foundation for the subsequent study of neutron star structure.

\subsection{NJL Model Formalism}
    This research primarily focuses on the two-flavor ($N_f=2$) NJL model, which includes up (u) and down (d) quarks. Its Lagrangian can be written as:
    \begin{equation}
        \mathcal{L} = \overline{\psi}(i\gamma^\mu \partial_\mu - \hat{m})\psi + \mathcal{L}_{\text{int}} \label{eq:lagrangian_total}
    \end{equation}
    Here, $\psi$ represents the quark field, $\gamma^\mu$ are the Dirac matrices, and $\hat{m}$ is the current quark mass matrix. In standard configurations, the current masses for u and d quarks are considered equal, i.e., $m_u = m_d$ \autocite{Yuan_2022_InteractingQuarkMatter}.

    The interaction term $\mathcal{L}_{\text{int}}$ consists of a four-fermion contact interaction, with a structure designed to capture the key symmetries of Quantum Chromodynamics (QCD) in the low-energy regime. We primarily consider the following two important interaction channels \autocite{Buballa_2005_NJLReview, Yuan_2022_InteractingQuarkMatter}:
    \begin{enumerate}
        \item \textbf{Scalar-Pseudoscalar Channel:} This term describes the attractive force responsible for spontaneous chiral symmetry breaking and is intimately related to the formation of mesons (such as $\pi$ mesons).
        \begin{equation}
            \mathcal{L}_{\sigma}^{(4)} = G_S[(\overline{\psi}\psi)^2 + (\overline{\psi}i\gamma_5\vec{\tau}\psi)^2] \label{eq:lagrangian_sigma}
        \end{equation}
        In this equation, $G_S$ is the scalar coupling constant, and $\vec{\tau}$ are the Pauli matrices acting in flavor space, representing the quark isospin degrees of freedom.
        \item \textbf{Vector Channel:} This term accounts for the short-range repulsive force between quarks, which significantly affects the pressure of quark matter and the structure of compact stars \autocite{Kojo_2015_PhenomenologicalQCD, Yuan_2022_InteractingQuarkMatter}.
        \begin{equation}
            \mathcal{L}_{V}^{(4)} = -G_V(\overline{\psi}\gamma^\mu\psi)^2 \label{eq:lagrangian_vector}
        \end{equation}
     Here, $G_V$ is the vector coupling constant. A positive $G_V$ signifies a repulsive interaction.
    \end{enumerate}
    Consequently, the total interaction Lagrangian is $\mathcal{L}_{\text{int}} = \mathcal{L}_{\sigma}^{(4)} + \mathcal{L}_{V}^{(4)}$. This specific form of the NJL model is capable of capturing several non-perturbative features of low-energy QCD, including chiral symmetry breaking and the dynamical generation of quark masses \autocite{Buballa_2005_NJLReview}.

    A central physical mechanism of the NJL model is the spontaneous breaking of chiral symmetry. This process allows quarks to acquire a substantial dynamical mass from their tiny current quark masses, which helps to explain why the quarks that form hadrons (like nucleons) appear to have a much larger effective mass \autocite{Nambu_1961, Hatsuda_1994_QCDPhenomenology, Buballa_2005_NJLReview}. Within the Mean-Field Approximation (MFA), the effective quark mass, $M$ (also known as the constituent quark mass), is self-consistently determined by the following chiral (or "gap") equation \autocite{Buballa_2005_NJLReview}:
    \begin{equation}
        M_f = m_f - 2G_S \langle \overline{\psi}_f \psi_f \rangle \label{eq:gap_equation}
    \end{equation}
    Here, $m_f$ is the current quark mass for flavor $f$, and $\langle \overline{\psi}_f \psi_f \rangle$ is the expected value of the quark condensate for that flavor. Both the quark condensate $\langle \overline{\psi}_f \psi_f \rangle$ and the quark number density $\rho_f$ depend on the temperature $T$ and the effective quark chemical potential $\mu_f^*$ \autocite{Buballa_2005_NJLReview}. The effective chemical potential $\mu_f^*$ accounts for the interaction between quarks and their environment, and it differs from the physical chemical potential $\mu_f$, especially when vector interactions are present \autocite{Yuan_2022_InteractingQuarkMatter}.

    For a two-flavor NJL model, the expression for the quark condensate typically involves an integral over momentum space. Due to the non-renormalizable nature of the model, a cutoff parameter $\Lambda$ (such as a three-momentum cutoff or Proper-Time Regularization) is necessary to handle divergent integrals \autocite{Buballa_2005_NJLReview}. At finite temperature and chemical potential, the gap equation includes Fermi-Dirac distribution functions to account for medium effects:
    \begin{equation}
        M_f = m_f + 4N_c G_S \int \frac{d^3 p}{(2\pi)^3} \frac{M_f}{E_p} (1 - n_p(E_p, \mu_f^*) - \bar{n}_p(E_p, \mu_f^*)) \label{eq:gap_equation_finite_T}
    \end{equation}
    where $E_p = \sqrt{\vec{p}^2 + M_f^2}$ is the quark energy, and $n_p$ and $\bar{n}_p$ are the Fermi-Dirac distribution functions for quarks and antiquarks, respectively \autocite{Buballa_2005_NJLReview}. At zero temperature, these distribution functions simplify to step functions.

    The introduction of the vector interaction term $\mathcal{L}_{V}^{(4)} = -G_V(\overline{\psi}\gamma^\mu\psi)^2$ establishes a clear link between the physical quark chemical potential $\mu_f$ and its effective chemical potential $\mu_f^*$. The effective chemical potential $\mu_f^*$ incorporates the mean-field interaction effects between quarks and the vector meson field, with the following expression \autocite{Buballa_2005_NJLReview, Yuan_2022_InteractingQuarkMatter}:
    \begin{equation}
        \mu_f^* = \mu_f - 2G_V \rho_f \label{eq:effective_chemical_potential}
    \end{equation}
    Here, $\mu_f$ is the physical chemical potential for quark flavor $f$, $G_V$ is the vector coupling constant, and $\rho_f = \langle \psi_f^\dagger \psi_f \rangle$ is the corresponding quark number density. \textbf{The number density $\rho_f$ for quarks at zero temperature is given by:}
    \begin{equation}
        \rho_f = \frac{N_c}{\pi^2} \int_0^{p_{F,f}} p^2 dp = \frac{N_c p_{F,f}^3}{3\pi^2} \quad \text{where } p_{F,f} = \sqrt{(\mu_f^*)^2 - M_f^2} \label{eq:quark_density_expression}
    \end{equation}
    \textbf{Here, $N_c=3$ is the number of colors, and $p_{F,f}$ is the Fermi momentum. When $\mu_f^* < M_f$, the Fermi momentum is 0, and the number density $\rho_f$ is also 0.} This relationship highlights that the presence of vector interactions means the actual energy state of quarks in the medium (described by the effective chemical potential) differs from the externally applied physical chemical potential. This interaction, which is generally repulsive, "offsets" a portion of the physical chemical potential, requiring a higher physical chemical potential to reach the same effective chemical potential state for the quarks.

\subsection{Charge Neutrality and Beta-Equilibrium Conditions}
    Matter within a neutron star must satisfy specific equilibrium conditions to remain stable under its extreme conditions. For quark matter, two fundamental conservation laws are charge neutrality and beta-equilibrium \autocite{Buballa_2005_NJLReview, Li_2022_HybridStarsQuarkCore}. These conditions impose strict constraints on the relationship between quark flavors and chemical potentials, which in turn profoundly impacts the quark matter EOS and the macroscopic properties of neutron stars.

    \textbf{1. Charge Neutrality Condition:}
    Given the nature of the strong interaction, quark matter must maintain overall charge neutrality to prevent the accumulation of immense Coulomb energy \autocite{Buballa_2005_NJLReview}. This requires that the total charge density from quarks and leptons (such as electrons) must be zero. For a two-flavor (u and d) quark system, the charge neutrality condition is expressed as \autocite{Li_2022_HybridStarsQuarkCore}:
    \begin{equation}
        \frac{2}{3}\rho_u - \frac{1}{3}\rho_d - \rho_e = 0 \label{eq:charge_neutrality}
    \end{equation}
    where $\rho_u$ and $\rho_d$ are the number densities of up and down quarks, respectively, and $\rho_e$ is the number density of electrons. The contribution of electrons, as leptons, cannot be neglected. The electron number density at zero temperature is determined by its chemical potential $\mu_e$ \autocite{Li_2022_HybridStarsQuarkCore}:
    \begin{equation}
        \rho_e(\mu_e) = \frac{\mu_e^3}{3\pi^2} \label{eq:electron_density}
    \end{equation}

    \textbf{2. Beta-Equilibrium Condition:}
    After a neutron star is formed, its internal matter achieves thermodynamic equilibrium through weak interaction processes, referred to as beta-equilibrium. These processes involve the interconversion of quarks, as well as quarks and leptons. For two-flavor (u, d) quark matter, the primary weak interaction processes are \autocite{Li_2022_HybridStarsQuarkCore}:
    \begin{align}
        d &\leftrightarrow u + e^- + \bar{\nu}_e \label{eq:beta_decay_d_u} \\
        u + e^- &\leftrightarrow d + \nu_e \label{eq:beta_decay_u_d}
    \end{align}
    Assuming neutrinos ($\nu_e$) are not trapped and can freely escape the star (which is the case for an old, cooled neutron star), these weak interactions lead to the following relationship between the chemical potentials of quarks and electrons \autocite{Li_2022_HybridStarsQuarkCore}:
    \begin{equation}
        \mu_d = \mu_u + \mu_e \label{eq:beta_equilibrium}
    \end{equation}
    When the charge neutrality condition (\ref{eq:charge_neutrality}) is combined with the beta-equilibrium condition (\ref{eq:beta_equilibrium}), only one independent variable remains among the quark chemical potentials. We typically choose the up-quark chemical potential $\mu_u$ as this variable and derive the expressions for $\mu_d$ and $\mu_e$ from these relationships. For instance, the down-quark chemical potential is $\mu_d = \mu_u + \mu_e$, where the value of $\mu_e$ is determined self-consistently by the charge neutrality condition \autocite{Buballa_2005_NJLReview, Li_2022_HybridStarsQuarkCore}.

    When calculating the quark matter EOS, these conditions must be satisfied simultaneously. This means the number densities of quarks ($\rho_u, \rho_d$) and electrons ($\rho_e$) are interconnected, collectively determining the system's pressure and energy density. This self-consistent calculation is a critical step in understanding the complex phase structure of matter within neutron stars.

\subsection{Thermodynamic Potential and Equation of State Calculation}
    \label{subsec:thermodynamic_potential}
    At zero temperature ($T=0$) and finite chemical potential, the macroscopic properties of quark matter are determined by its Grand Canonical Potential, also known as the thermodynamic potential $\Omega$ \autocite{Buballa_2005_NJLReview, Yuan_2022_InteractingQuarkMatter}. By integrating over the quark energy spectrum, we can derive the expression for this potential. In the mean-field approximation, the total thermodynamic potential of the system includes contributions from both quarks and electrons \autocite{Buballa_2005_NJLReview}.

    The general form of the total thermodynamic potential $\Omega(T, \mu; M, \tilde{\mu})$ is given by \autocite{Buballa_2005_NJLReview}:
    \begin{equation}
        \Omega(T,\mu;M,\tilde{\mu}) = \Omega_M(T,\tilde{\mu}) + \frac{(M-m)^2}{4G_S} - \frac{(\mu-\tilde{\mu})^2}{4G_V} + \text{const.} \label{eq:omega_total}
    \end{equation}
    \textbf{Here, $\Omega_M(T=0, \tilde{\mu})$ represents the contribution from a free Fermi gas (quarks and antiquarks) at zero temperature. For two-flavor quarks ($N_f=2$), its expression is:}
    \begin{equation}
        \Omega_M(T=0, \tilde{\mu}) = -\frac{N_c}{24\pi^2} \sum_{f=u,d} \left[ \tilde{\mu}_f \sqrt{\tilde{\mu}_f^2 - M_f^2} (2\tilde{\mu}_f^2 - 5M_f^2) + 3M_f^4 \ln\left(\frac{\tilde{\mu}_f + \sqrt{\tilde{\mu}_f^2 - M_f^2}}{M_f}\right) \right] \label{eq:omega_M_quark_expression}
    \end{equation}
    \textbf{where the summation $\sum_{f=u,d}$ runs over up and down quarks, and $N_c=3$ is the number of colors.}
    
    The self-consistent equations are then obtained by minimizing the thermodynamic potential with respect to its auxiliary variables (such as $M$ and $\tilde{\mu}$) \autocite{Buballa_2005_NJLReview}. Once a stable self-consistent solution is found, fundamental EOS quantities like pressure $P$ and energy density $\epsilon$ can be derived using standard thermodynamic relations \autocite{Buballa_2005_NJLReview, Yuan_2022_InteractingQuarkMatter}:
    \begin{align}
        P &= -\Omega \label{eq:pressure_omega} \\
        \epsilon &= \sum_f \mu_f \rho_f - P \label{eq:energy_density_thermo_quark}
    \end{align}
    Here, $\mu_f$ and $\rho_f$ are the chemical potential and particle number density for quark flavor $f$. The summation $\sum_f$ includes all existing quark flavors, which in this study are primarily the u and d quarks. For neutron star matter, the contribution of leptons (e.g., electrons) must also be considered, so the energy density is more accurately expressed as $\epsilon = \sum_i \mu_i \rho_i - P$, where $i$ runs over all constituent particles (quarks and leptons). The baryon number density $\rho_B$ is given by the derivative of pressure with respect to the baryon chemical potential: $\rho_B = \frac{\partial P}{\partial \mu_B}$.
    
\section{Hadronic and Hybrid Equation of State Construction}
\label{sec:eos-construction}
    A central challenge in understanding the internal structure and evolution of neutron stars is the accurate construction of an Equation of State (EOS) that describes matter under extreme conditions. The density within a neutron star varies drastically, from the relatively low densities in its crust to ultra-high densities in the core that can far exceed the nuclear saturation density. No single physical model can comprehensively cover such a wide range. Therefore, this study employs a method of layered construction and smooth interpolation to combine hadronic and quark EOSs, with the aim of creating a thermodynamically consistent hybrid EOS across the entire density range.

\subsection{Hadronic Equation of State Selection}
\label{subsec:hadronic-eos}
    The outer regions of a neutron star are composed of nuclear matter, and its EOS determines the physical properties of the stellar crust and outer core. 
    
    In RMF theory, the Lagrangian density for hadronic matter typically includes a nucleonic component, meson self-interaction components, and mixed-interaction terms between mesons \autocite{Agrawal_2010_ERMF_neutron_skin_original, Fortin_2016_BSR_EOS}. A general RMF Lagrangian density, which forms the basis for such parameterizations, can be expressed as:
    \begin{equation}
        \mathcal{L}=\mathcal{L}_{NM}+\mathcal{L}_{\sigma}+\mathcal{L}_{\omega}+\mathcal{L}_{\rho}+\mathcal{L}_{\sigma\omega\rho} \label{eq:hadronic_lagrangian}
    \end{equation}
    where:
    \begin{itemize}
        \item $\mathcal{L}_{NM}$ is the nucleonic part of the Lagrangian, which describes the free motion of nucleons (neutrons $n$ and protons $p$) and their coupling to the meson fields:
        \begin{equation*}
            \mathcal{L}_{NM}=\sum_{H=n,p}\overline{\psi}_{H}[i\gamma^{\mu}\partial_{\mu}-(M-g_{\sigma}\sigma)-(g_{\omega}\gamma^{\mu}\omega_{\mu}+\frac{1}{2}g_{\rho}\gamma^{\mu}\vec{\tau}\cdot\vec{\rho}_{\mu})]\psi_{H}
        \end{equation*}
        Here, $\psi_H$ represents the nucleon field, $M$ is the nucleon mass, $g_{\sigma}, g_{\omega}, g_{\rho}$ are the coupling constants for the nucleons to the $\sigma, \omega, \rho$ meson fields, respectively, and $\vec{\tau}$ is the isospin matrix.
        \item $\mathcal{L}_{\sigma}$, $\mathcal{L}_{\omega}$, and $\mathcal{L}_{\rho}$ describe the dynamics and self-interaction terms of the $\sigma$ meson (scalar-isoscalar), $\omega$ meson (vector-isoscalar), and $\rho$ meson (vector-isovector) fields. For example, the $\sigma$ meson term includes a mass term and nonlinear self-coupling terms:
        \begin{equation*}
            \mathcal{L}_{\sigma}=\frac{1}{2}(\partial^{\mu}\sigma\partial_{\mu}\sigma-m_{\sigma}^{2}\sigma^{2})-\frac{\kappa_{3}}{6M}g_{\sigma}^{3}m_{\sigma}^{2}\sigma^{3}-\frac{\kappa_{4}}{24M^{2}}g_{\sigma}^{4}m_{\sigma}^{2}\sigma^{4}
        \end{equation*}
        where $m_{\sigma}$ is the $\sigma$ meson mass, and $\kappa_3, \kappa_4$ are nonlinear coupling coefficients. \textbf{For the $\omega$ meson term $\mathcal{L}_{\omega}$, its form typically includes:}
        \begin{equation*}
            \mathcal{L}_{\omega} = -\frac{1}{4}F_{\omega}^{\mu\nu}F_{\omega, \mu\nu} + \frac{1}{2}m_{\omega}^{2}\omega^{\mu}\omega_{\mu} + \frac{\zeta_0}{4!} (g_{\omega}\omega^{\mu}\omega_{\mu})^2
        \end{equation*}
        \textbf{Here, $F_{\omega}^{\mu\nu} = \partial^{\mu}\omega^{\nu} - \partial^{\nu}\omega^{\mu}$ is the field strength tensor for the $\omega$ meson, $m_{\omega}$ is the $\omega$ meson mass, and $\zeta_0$ is its nonlinear self-coupling coefficient.}
        
        \textbf{For the $\rho$ meson term $\mathcal{L}_{\rho}$, its form typically includes:}
        \begin{equation*}
            \mathcal{L}_{\rho} = -\frac{1}{4}\vec{F}_{\rho}^{\mu\nu}\cdot\vec{F}_{\rho, \mu\nu} + \frac{1}{2}m_{\rho}^{2}\vec{\rho}^{\mu}\cdot\vec{\rho}_{\mu} + \frac{\xi_0}{4!} (g_{\rho}\vec{\rho}^{\mu}\cdot\vec{\rho}_{\mu})^2
        \end{equation*}
        \textbf{Here, $\vec{F}_{\rho}^{\mu\nu} = \partial^{\mu}\vec{\rho}^{\nu} - \partial^{\nu}\vec{\rho}^{\mu} - g_{\rho}(\vec{\rho}^{\mu}\times\vec{\rho}^{\nu})$ is the field strength tensor for the $\rho$ meson, $m_{\rho}$ is the $\rho$ meson mass, and $\xi_0$ is its nonlinear self-coupling coefficient. These self-interaction terms are vital for describing the saturation behavior of mesons at high densities, particularly for accurately characterizing the properties of asymmetric nuclear matter \autocite{Agrawal_2010_ERMF_neutron_skin_original, Fortin_2016_BSR_EOS, Reinhard_1989_RMFReview}.}
        \item $\mathcal{L}_{\sigma\omega\rho}$ describes the mixed interaction terms between the meson fields, which are crucial for precisely characterizing nuclear matter properties (especially the density dependence of the symmetry energy) \autocite{Agrawal_2010_ERMF_neutron_skin_original}:
        \begin{align*}
            \mathcal{L}_{\sigma\omega\rho} &= \frac{\eta_{1}}{2M}g_{\sigma}m_{\omega}^{2}\sigma\omega^{\mu}\omega_{\mu} + \frac{\eta_{2}}{4M^{2}}g_{\sigma}^{2}m_{\omega}^{2}\sigma^{2}\omega^{\mu}\omega_{\mu} + \frac{\eta_{3}}{2M}g_{\sigma}m_{\rho}^{2}\sigma\rho^{\mu}\rho_{\mu} \\
            &+ \frac{\eta_{4}}{4M^{2}}g_{\sigma}^{2}m_{\rho}^{2}\sigma^{2}\rho^{\mu}\rho_{\mu} + \frac{\eta_{5}}{4M^{2}}g_{\omega}^{2}m_{\rho}^{2}\omega^{\mu}\omega_{\mu}\rho^{\mu}\rho_{\mu}
        \end{align*}
        These $\eta$ coefficients are phenomenologically determined in the RMF model by fitting the ground-state properties of finite nuclei and the nuclear matter parameters at saturation density to optimize the model.
    \end{itemize}

    To accurately describe the nuclear matter across the entire star, from the crust to the core, we select the \textbf{DDME2 model} \autocite{Lalazissis_2005_DDME2}. While standard nonlinear RMF models employ constant couplings ($g_{\sigma}, g_{\omega}, g_{\rho}$) and achieve saturation via nonlinear meson self-interaction terms (e.g., $\kappa_3, \kappa_4, \zeta_0$), density-dependent models like DDME2 adopt a different approach. In this framework, the nonlinear self-couplings are typically omitted ($\kappa_3=\kappa_4=\zeta_0=0$), and the medium effects are instead encoded by making the couplings $g_{\sigma}(\rho_B), g_{\omega}(\rho_B), g_{\rho}(\rho_B)$ explicit functions of the local baryon density $\rho_B$ \autocite{Lalazissis_2005_DDME2}. This allows for a more realistic description of nuclear matter properties away from saturation.

    The DDME2 parameterization is well-calibrated, predicting nuclear saturation properties such as a saturation density $n_0=0.152 \text{ fm}^{-3}$, binding energy $E/A=-16.14 \text{ MeV}$, incompressibility $K=251 \text{ MeV}$, and symmetry energy $J=32.3 \text{ MeV}$ \autocite{Lalazissis_2005_DDME2}, all of which are consistent with experimental data. A key advantage of this model is that it provides a \textbf{unified equation of state} that includes both the inner and outer crust, as well as the liquid core. This approach eliminates the need for a separate low-density crust EOS (like the BPS model \autocite{Baym_1971_BPS}) and the associated stitching procedures discussed in works like \autocite{Fortin_2016_BSR_EOS}, ensuring thermodynamic consistency across the entire hadronic phase.
    
    It is important to note that RMF models do not typically include explicit hyperon degrees of freedom. This is because the interactions between hyperons and nucleons, as well as hyperon-hyperon interactions, are still subject to considerable uncertainty, and most models predict that hyperons begin to appear at densities around $n_B \sim 2-3n_0$, a range that coincides with the hadron-quark crossover region. Therefore, their impact on the EOS must be considered as part of the hybrid state construction.

    To describe the neutron star matter within the RMF framework, we explicitly impose the conditions of $\beta$-equilibrium and charge neutrality, analogous to the treatment in the quark sector. The chemical equilibrium is maintained via weak interactions, leading to the relation between chemical potentials: $\mu_n = \mu_p + \mu_e$. Simultaneously, the charge neutrality condition requires the balance of proton and electron number densities, $\rho_p = \rho_e$. By solving the coupled equations of motion for meson fields subject to these constraints, the composition of hadronic matter is self-consistently determined for any given baryon density.
    
\subsection{Hadron-Quark Hybrid Equation of State Construction}
    The extreme density environment in the core of a neutron star may induce a deconfinement phase transition of hadronic matter, leading to the formation of quark matter composed of free quarks and gluons. This gives rise to the possibility of "hybrid stars" \autocite{Baym_2018_FromHadronsToQuarks, Li_2022_HybridStarsQuarkCore, Li_2022_PropertiesOfHybridStars}. This study employs a "three-window" approach \autocite{Kojo_2015_PhenomenologicalQCD, Baym_2018_FromHadronsToQuarks} to construct the hadron-quark hybrid EOS. This method allows for a smooth transition, or "crossover," region between the hadronic and quark phases.

    This approach avoids potential issues like pressure discontinuities or unphysical sound speeds associated with first-order phase transitions and better represents a continuous evolution from hadrons to quarks \autocite{Kojo_2015_PhenomenologicalQCD, Baym_2018_FromHadronsToQuarks}. We divide the EOS into three regions based on the baryon chemical potential $\mu_B$:
    \begin{enumerate}
        \item \textbf{Low-Density Hadronic Region ($\mu_B < \mu_{BL}$):} In this region, matter is described by the unified DDME2 hadronic EOS. $\mu_{BL}$ defines the \textbf{validity boundary of the pure hadronic description}, typically corresponding to a baryon number density of $n_B \sim (1-2)n_0$. Above this density, the \textbf{finite-size effects of nucleons become significant}, and the reliability of point-particle hadronic models diminishes due to the onset of quark percolation \autocite{Baym_2018_FromHadronsToQuarks, Kojo_2015_PhenomenologicalQCD}.
        \item \textbf{High-Density Quark Region ($\mu_B > \mu_{BU}$):} In this region, matter is assumed to be fully deconfined quark matter. We use the two-flavor NJL model as described in Section 2, which incorporates quark vector interactions and chiral symmetry breaking effects. $\mu_{BU}$ is the defined lower boundary chemical potential for the quark phase, generally corresponding to a baryon number density of $n_B \sim (4-7)n_0$. Below this density, quark confinement effects become significant, limiting the applicability of a quark model \autocite{Baym_2018_FromHadronsToQuarks, Kojo_2015_PhenomenologicalQCD}.
        \item \textbf{Intermediate Transition Region ($\mu_{BL} \le \mu_B \le \mu_{BU}$):} This is a hadron-quark mixed phase or a continuous crossover region, which is difficult to calculate precisely from first principles. This study uses a phenomenological interpolation method to describe the EOS in this region. We use a quintic polynomial $\mathcal{P}(\mu_B) = \sum_{m=0}^{5}C_{m}\mu_{B}^{m}$ to connect the pressure-baryon chemical potential relationship of the hadronic and quark phases (as detailed in Appendix \ref{appendix:interpolation_coeffs}) \autocite{Li_2022_HybridStarsQuarkCore, Baym_2018_FromHadronsToQuarks, Kojo_2015_PhenomenologicalQCD}. The polynomial coefficients $C_m$ are determined by applying the following boundary conditions, which ensure the thermodynamic consistency and smoothness of the entire hybrid EOS:
        \begin{itemize}
            \item At $\mu_B = \mu_{BL}$, the pressure $P(\mu_B)$ and its first two derivatives with respect to $\mu_B$ (i.e., the baryon number density $\rho_B(\mu_B)$ and the baryon number susceptibility $\partial \rho_B / \partial \mu_B$) must match those of the hadronic EOS.
            \item At $\mu_B = \mu_{BU}$, the pressure $P(\mu_B)$ and its first two derivatives with respect to $\mu_B$ must match those of the quark EOS.
        \end{itemize}
        These conditions ensure the continuity of pressure, number density, and number susceptibility in the transition region, thereby avoiding unphysical jumps or unstable areas \autocite{Fortin_2016_BSR_EOS}.
    \end{enumerate}
    This piecewise construction and interpolation method yields a hybrid EOS, $P(\mu_B)$, that smoothly transitions across the entire density range and satisfies fundamental physical constraints \autocite{Baym_2018_FromHadronsToQuarks, Kojo_2015_PhenomenologicalQCD}:
    \begin{itemize}
        \item \textbf{Pressure Continuity:} The pressure $P$ must be a continuous function of the baryon chemical potential $\mu_B$.
        \item \textbf{Thermodynamic Stability:} The baryon number density $\rho_B = \partial P / \partial \mu_B$ must be a monotonically increasing function of $\mu_B$, i.e., $\partial^2 P / \partial \mu_B^2 > 0$, to ensure the system's stability against density fluctuations.
        \item \textbf{Causality:} The speed of sound squared, $v_s^2 = \partial P / \partial \epsilon$, in the medium must be less than or equal to the speed of light squared, $c^2$, i.e., $v_s^2/c^2 \le 1$. This is a fundamental physical requirement that signals cannot propagate faster than light.
    \end{itemize}
    A hybrid EOS that satisfies these constraints will provide a reliable physical input for subsequent calculations of neutron star structure and properties.

\subsection{Model Parameters and Consistency}
\label{subsec:njl-parameters}
    The parameter selection for the NJL model is designed to ensure consistency with the density scale of the DDME2 hadronic phase. Specifically, we explicitly adopt the nuclear saturation density predicted by DDME2 ($n_0 \approx 0.152 \text{ fm}^{-3}$) as the fundamental reference unit for defining the crossover boundaries ($BL$ and $BU$).

    For the quark sector, we adopt a standard parameterization scheme for the NJL model. Based on the literature \autocite{Buballa_2005_NJLReview}, we adopted one of their parameter sets and fine-tuned it to obtain the following parameters. The current quark mass is set to $m_u = m_d = 5.50$ MeV, and the three-momentum cutoff is fixed at $\Lambda = 664.3$ MeV. Regarding the scalar coupling strength, we treat the dimensionless factor $G_S\Lambda^2$ as a tunable parameter in this study to explore the stiffness of the EOS at high densities. As detailed in Section \ref{sec:glambda2-impact}, we establish a fiducial benchmark value of $G_S\Lambda^2 = 1.970$ (corresponding to a constituent quark mass $M \approx 335$ MeV in vacuum), which best satisfies current observational constraints and the causality constraint required for constructing a hybrid star EOS. The vector coupling ratio is subsequently set to a fiducial value of $G_V/G_S = 0.23$. These values for parameters are calibrated to reproduce the vacuum properties of the pion: the pion decay constant $f_\pi = 92.4$ MeV and the pion mass $m_\pi = 135.0$ MeV.

    The phase transition window for this benchmark is set by $BL=1.0$ and $BU=7.60$. This early transition onset ($BL=1.0$) is primarily driven by recent stringent observational constraints. Specifically, the compact radius measurement of PSR J0437-4715 \autocite{Bogdanov_2019_J0437} necessitates a softer EOS at intermediate densities. Our extensive parameter space exploration indicates that models with a delayed transition onset ($BL \ge 1.2$) fail to simultaneously satisfy the mass and radius constraints imposed by massive pulsars and NICER observations. Consequently, a transition initiating near the nuclear saturation density is identified as a necessary feature to reconcile these multi-messenger data. 
    
    It is important to note that in our crossover construction, setting $BL \approx 1.0$ does not imply an abrupt termination of the hadronic phase or the immediate presence of bulk quark matter at saturation. Instead, physically, this choice marks the \textbf{limit of validity for the point-particle approximation} of nucleons. Theoretical studies on quarkyonic matter and quark percolation suggest that at densities exceeding $n_0$, the phase space may already be partially populated by quark states or that nucleon wave functions significantly overlap, leading to a breakdown of the pure hadronic description~\autocite{Baym_2018_FromHadronsToQuarks, Kojo_2015_PhenomenologicalQCD}. Our phenomenological results support this picture, indicating an \textbf{early percolation of quark degrees of freedom} and the onset of precursor deconfinement effects. These effects naturally soften the EOS compared to pure RMF predictions. Therefore, initiating the crossover near $n_0$ serves as an \textbf{effective description} to capture this physical softening---necessitated by the compact radius data of PSR J0437-4715---without explicitly modeling the complex many-body dynamics of nucleon structure breakdown.
    
\label{sec:tov-equation}
    The macroscopic properties of a neutron star-such as mass, radius, and deformability under an external gravitational field, characterized by the tidal deformability parameter-are directly governed by the Equation of State (EOS) of its internal matter. To derive these macroscopic characteristics from the microscopic EOS, we must solve the stellar structure equations, typically the Tolman–Oppenheimer–Volkoff (TOV) equations---within the framework of general relativity and account for physical processes such as tidal deformation.

\subsection{Tolman-Oppenheimer-Volkoff (TOV) Equations and Their Numerical Solution}
\label{subsec:tov-numerical-solution}
    For a static, spherically symmetric neutron star, its internal structure is described by the \textbf{Tolman-Oppenheimer-Volkoff (TOV) equations}\autocite{Tolman_1939_StaticSolutions, Oppenheimer_1939_MassiveNeutronCores}. This set of equations is a simplified form of Einstein's field equations for a spherically symmetric fluid distribution, which precisely describes how the internal pressure $P(r)$ and enclosed mass $M(r)$ change with the radial coordinate $r$:
    \begin{align}
        \frac{dP}{dr} &= -\frac{G(\epsilon + P/c^2)(M(r) + 4\pi r^3 P/c^2)}{r(r - 2GM(r)/c^2)} \label{eq:tov_pressure} \\
        \frac{dM}{dr} &= 4\pi r^2 \epsilon \label{eq:tov_mass}
    \end{align}
    Here, $G$ is the gravitational constant, $c$ is the speed of light in a vacuum, and $\epsilon$ is the energy density, which is closely related to the pressure $P$ through the EOS, $\epsilon(P)$, constructed previously.

    To numerically solve the TOV equations, we typically use the \textbf{Runge-Kutta method} for integration. The process begins at the center of the star ($r=0$), where an initial central pressure $P_c$ and an initial mass of $M(0)=0$ are set. As the radial distance $r$ increases, the pressure $P(r)$ gradually decreases. The integration stops when the pressure falls to a preset surface threshold. At this point, the radial distance $r$ is the star's radius $R$, and $M(R)$ is the total mass $M$ of the neutron star corresponding to the central pressure $P_c$. By systematically scanning a range of different central pressures $P_c$, we can calculate multiple sets of $(R, M)$ values, which allows us to plot the complete \textbf{mass-radius ($M-R$) relationship curve} for the neutron star. This curve provides a direct visual representation of how the EOS influences the star's macroscopic structure and can be directly compared with astronomical observational data \autocite{Lattimer_2016_EOSConstraints, Haensel_2007_NeutronStarsBook}.

    The key observational constraints we use for comparison are:
    \begin{itemize}
        \item \textbf{PSR J0740+6620}: A known massive pulsar with a mass of approximately $M = 2.08 \pm 0.07 M_\odot$ and a radius of about $12.49^{+1.28}_{-0.88}$ km \autocite{Salmi_2024_PSRJ0740, Cromartie_2019}.
        \item \textbf{PSR J0030+0451}: NICER observations provide a joint constraint on its mass and radius, $M = 1.34^{+0.15}_{-0.16} M_\odot$ and $R = 12.71^{+1.14}_{-1.19}$ km \autocite{Riley_2019_PSRJ0030}.
        \item \textbf{PSR J0437-4715}: Observations of this pulsar also constrain its mass and radius to approximately $M = 1.44 \pm 0.07 M_\odot$ and about $11.36^{+0.95}_{-0.63}$ km \autocite{Bogdanov_2019_J0437}.
    \end{itemize}

\subsection{Tidal Deformability $\boldsymbol{\Lambda}$ and Neutron Star Macroscopic Properties}
\label{subsec:tidal-deformability}
    The \textbf{tidal deformability parameter $\boldsymbol{\Lambda}$} is a key physical quantity that measures how much a neutron star deforms under an external gravitational field (for example, during a binary neutron star merger event) \autocite{Hinderer_2008_TidalLoveNumbers, Damour_Nagar_2009_RelativisticTidalDeformability}. This parameter, by influencing the phase evolution of the gravitational wave signal, provides a unique and powerful way to precisely constrain the EOS of dense matter \autocite{Abbott_2017_GW170817, Read_2009_EOSGW, Paschalidis_2017_NeutronStarsReview}.

    In the framework of general relativity, for a static, spherically symmetric, non-rotating star, when it is perturbed by an external \textbf{quadrupolar tidal field} $\mathcal{E}_{ij}$, the star induces its own quadrupolar moment $Q_{ij}$. The \textbf{Love number} $k_l$ (typically referring to the quadrupolar deformation, $k_2$) is defined as the dimensionless proportionality constant that links this induced quadrupolar moment to the external tidal field \autocite{Love_1912_Tides, Binnington_2009_TidalLoveNumbers, Flanagan_2008_ConstrainingLoveNumbers}. The tidal deformability parameter $\Lambda$ further relates the Love number to the star's mass $M$ and radius $R$ with the following specific relationship \autocite{Flanagan_2008_ConstrainingLoveNumbers, Hinderer_2010_TidalDeformabilityRealisticEOS}:
    \begin{equation}
        \Lambda = \frac{2}{3} k_2 \left(\frac{c^2 R}{G M}\right)^5 \label{eq:lambda_k2_relation_new}
    \end{equation}
    where $R$ and $M$ are the neutron star's radius and mass, respectively, and $k_2$ is the quadrupolar Love number. Calculating $k_2$ is a critical step in determining $\Lambda$.

    The calculation of $k_2$ involves solving the linearized Einstein equations for the star when it is subjected to a small external tidal field. First, the unperturbed spherically symmetric stellar metric (obtained from the solution of the TOV equations) is linearly perturbed. In the Regge-Wheeler gauge, for a static, even-parity (electric-type) quadrupolar ($l=2$) perturbation, the metric perturbation can be described by a master function $H(r)$ \autocite{Hinderer_2008_TidalLoveNumbers, Thorne_1967_NonRadialPulsation}. The master function $H(r)$ satisfies a second-order ordinary differential equation:
    \begin{equation}
        r H^{\prime\prime}(r) - H^{\prime}(r) F(r) - H(r) Q(r) = 0 \label{eq:H_equation}
    \end{equation}
    Here, the prime denotes a derivative with respect to the radial coordinate $r$, and the coefficients $F(r)$ and $Q(r)$ are complex functions determined by the background spacetime (i.e., the solution of the TOV equations) (their explicit forms are given in Appendix \ref{appendix:tidal_functions}), depending on the pressure, energy density, and enclosed mass at $r$.

    Solving this equation typically involves a numerical integration method. Starting from the center of the star ($r \rightarrow 0$), the initial condition for the master function $H(r)$ is determined by the regularity requirement at the origin, which for $l=2$ is $H(r) = a_0 r^2$\autocite{Hinderer_2008_TidalLoveNumbers, Damour_Nagar_2009_RelativisticTidalDeformability}, where $a_0$ is an arbitrary constant. The integration proceeds outward to the stellar surface at $r=R$. By matching the internal numerical solution $H(R)$ and its radial derivative $H'(R)$ with the external analytical solution at the stellar surface $r=R$, we can ultimately calculate the Love number $k_2$. Specifically, the Love number $k_2$ is determined by matching the internal and external solutions at the stellar surface. It relies on the star's compactness parameter $C=GM/(Rc^2)$ and the surface logarithmic derivative $y$ of the master function, defined as $y = R H'(R)/H(R)$. The explicit expression derived by Hinderer et al. is given by \autocite{Hinderer_2008_TidalLoveNumbers, Hinderer_2010_TidalDeformabilityRealisticEOS}:
    \begin{align}
        k_2 &= \frac{8C^5}{5}(1-2C)^2[2+2C(y-1)-y] \nonumber \\
        &\times \Big\{2C[6-3y+3C(5y-8)] \nonumber \\
        &+ 4C^3[13-11y+C(3y-2)+2C^2(1+y)] \nonumber \\
        &+ 3(1-2C)^2[2-y+2C(y-1)]\ln(1-2C)\Big\}^{-1} \label{eq:k2_explicit}
    \end{align}
    This formula connects the macroscopic stiffness of the EOS directly to the observable gravitational wave signature.
    
    The Love number $k_2$ is highly sensitive to the "stiffness" of the EOS: typically, a stiffer EOS leads to a larger Love number, and vice versa \autocite{Hinderer_2010_TidalDeformabilityRealisticEOS, Annala_2018_EOSStronglyInteractingMatter}.

\section{Impact of the Vector Coupling Constant $G_V$ on Neutron Star Macroscopic Properties}
\label{sec:gv-impact}
    To understand the underlying physics that allows the model to succeed, we now turn to a more systematic investigation of how individual key parameters, such as $G_V$, regulate the macroscopic properties.
    In the NJL model, the vector coupling constant $G_V$ describes the repulsive interaction between quarks mediated by vector meson exchange. This interaction is particularly important at high densities, as it can effectively increase the "stiffness" of the equation of state, thereby having a decisive impact on macroscopic properties such as the maximum mass of a neutron star \autocite{Kojo_2015_PhenomenologicalQCD, Yuan_2022_InteractingQuarkMatter}. To systematically investigate the effect of this key parameter, we establish a baseline model and vary only the $G_V/G_S$ ratio while keeping all other parameters fixed.

\subsection{Parameter Sets and Comparison of Key Properties}
\label{subsec:gv-params-table}
    We utilize a \textbf{fiducial benchmark parameter set} selected from our parameter space exploration. This set is chosen because it simultaneously satisfies all key observational constraints (including the mass of PSR J0740+6620, the tidal deformability limit from GW170817, and NICER radius measurements). The parameters for this fiducial benchmark are: $G_S\Lambda^2 = 1.970$, $BL=1.0$, and $BU=7.60$.
    
    The baseline $G_V/G_S$ ratio for this set is 0.23. In the context of this section's analysis and figures (e.g., Figure \ref{fig:gv_comparison_combined_new}), this baseline set is referred to as \textbf{Set 3}. We then construct four additional sets by varying this $G_V/G_S$ ratio from 0.13 to 0.33, while keeping all other parameters ($G_S\Lambda^2$, $BL$, $BU$) fixed, to observe its impact.
    Table \ref{tab:gv_comparison_properties_new} quantitatively summarizes the key macroscopic properties for these five parameter sets.

\begin{table}[htbp]
    \centering
    \caption{Comparison of Key Neutron Star Macroscopic Properties for Different $G_V$ Parameters, based on the optimal benchmark (Set 3).
    \newline
    \textbf{Note:} These parameter sets exhibit a causality violation ($v_s^2/c^2 > 1$) in the phase transition region (see Figure \ref{fig:gv_comparison_combined_new}a); thus, their results are for theoretical trend reference only and are not physically acceptable.
    }
    \label{tab:gv_comparison_properties_new}
    \small
    \begin{tabular}{lccccc}
        \hline\hline
        \textbf{Parameter Set} & \textbf{$G_V/G_S$ Ratio} & \textbf{$M_{max}$ ($M_\odot$)} & \textbf{$R_{1.4}$ (km)} & \textbf{$\Lambda_{1.4}$} & \textbf{Status} \\
        \hline
        Set 1 & 0.13 & 2.11 & 12.17 & 376.0 & Success \\
        Set 2 & 0.18 & 2.18 & 12.19 & 381.3 & Success \\
        Set 3 & 0.23 & 2.25 & 12.20 & 383.3 & Success \\
        Set 4$^{a}$ & 0.27 & 2.30 & 12.20 & 383.7 & Violated \\
        Set 5$^{a}$ & 0.33 & 2.37 & 12.19 & 382.1 & Violated \\
        \hline\hline
    \end{tabular}
\end{table}

    The table clearly shows several critical trends. First, there is a strong, positive correlation between the value of $G_V$ and the neutron star's maximum mass ($M_{max}$). As $G_V/G_S$ increases from 0.13 to 0.33, $M_{max}$ systematically grows from $2.11 M_\odot$ to $2.37 M_\odot$. This directly confirms the stiffening effect of the repulsive vector interaction on the high-density equation of state. Notably, the maximum mass for our benchmark Set 3 ($2.25 M_{\odot}$) aligns remarkably well with the recent precise inference of $M_{TOV} = 2.25^{+0.08}_{-0.07} M_{\odot}$ derived from multi-messenger data \autocite{Fan_2024_MaxMass}.
    
    Second, a notable finding in this parameter regime is the relative insensitivity of $R_{1.4}$ and $\Lambda_{1.4}$ to variations in $G_V$.For all physically viable sets (1-3), $R_{1.4}$ remains nearly constant around $12.17-12.20$ km, and $\Lambda_{1.4}$ clusters tightly within the range of $376-383$.This suggests that while $G_V$ serves as the primary control parameter for tuning $M_{max}$, it exerts minimal influence on the properties of canonical $1.4 M_\odot$ stars within this model framework.
    
    Third, and most importantly, causality considerations impose a strict upper bound on $G_V$. As shown in the table and discussed in the next section, parameter sets with $G_V/G_S \ge 0.27$ (Sets 4 and 5) become acausal, rendering them physically invalid.

    Finally, it is worth noting that the strength of the vector coupling is not only relevant for the stiffness of the EOS in neutron stars but is also independently constrained by relativistic heavy-ion collision (HIC) experiments.Studies based on transport models incorporating NJL-based partonic mean fields have demonstrated that the vector interaction---being repulsive for quarks and attractive for antiquarks in a baryon-rich medium---is a crucial mechanism for explaining the observed elliptic flow ($v_2$) splitting between particles and antiparticles (e.g., $p$ and $\bar{p}$, $K^+$ and $K^-$) in the RHIC Beam Energy Scan data \autocite{Song_2014_EllipticFlow, Xu_2014_EllipticFlowSplitting}.Specifically, to quantitatively reproduce the measured relative $v_2$ differences, these phenomenological studies typically favor a relatively strong vector coupling, estimated in the range of $G_V/G_S \approx 0.5-1.1$ \autocite{Xu_2014_EllipticFlowSplitting}.By comparison, our fiducial choice of $G_V/G_S = 0.23$ (and the parameter exploration up to $0.33$) represents a more conservative estimate.This choice ensures that the hybrid EOS strictly adheres to causality limits at ultra-high densities, while still incorporating the essential repulsive physics supported by terrestrial HIC observations.

\subsection{Graphical Analysis of the EOS and Macroscopic Properties}
\label{subsec:gv-plots-analysis}
\begin{figure*}[htbp]
    \centering
    \begin{minipage}[b]{0.48\textwidth}
        \centering
        \includegraphics[width=\textwidth]{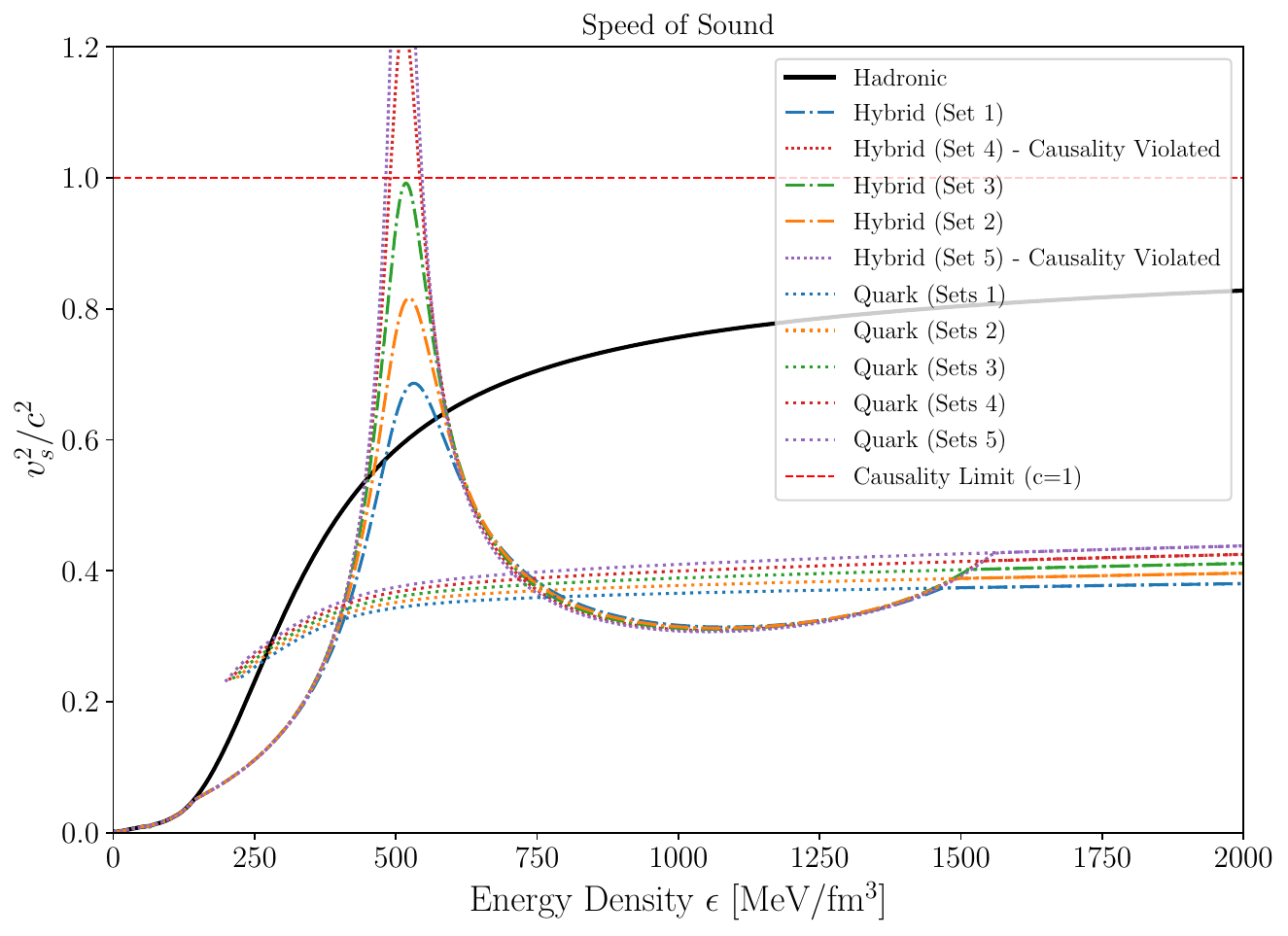}
        \centerline{(a) Speed of Sound vs. Energy Density}
    \end{minipage}
    \hfill
    \begin{minipage}[b]{0.48\textwidth}
        \centering
        \includegraphics[width=\textwidth]{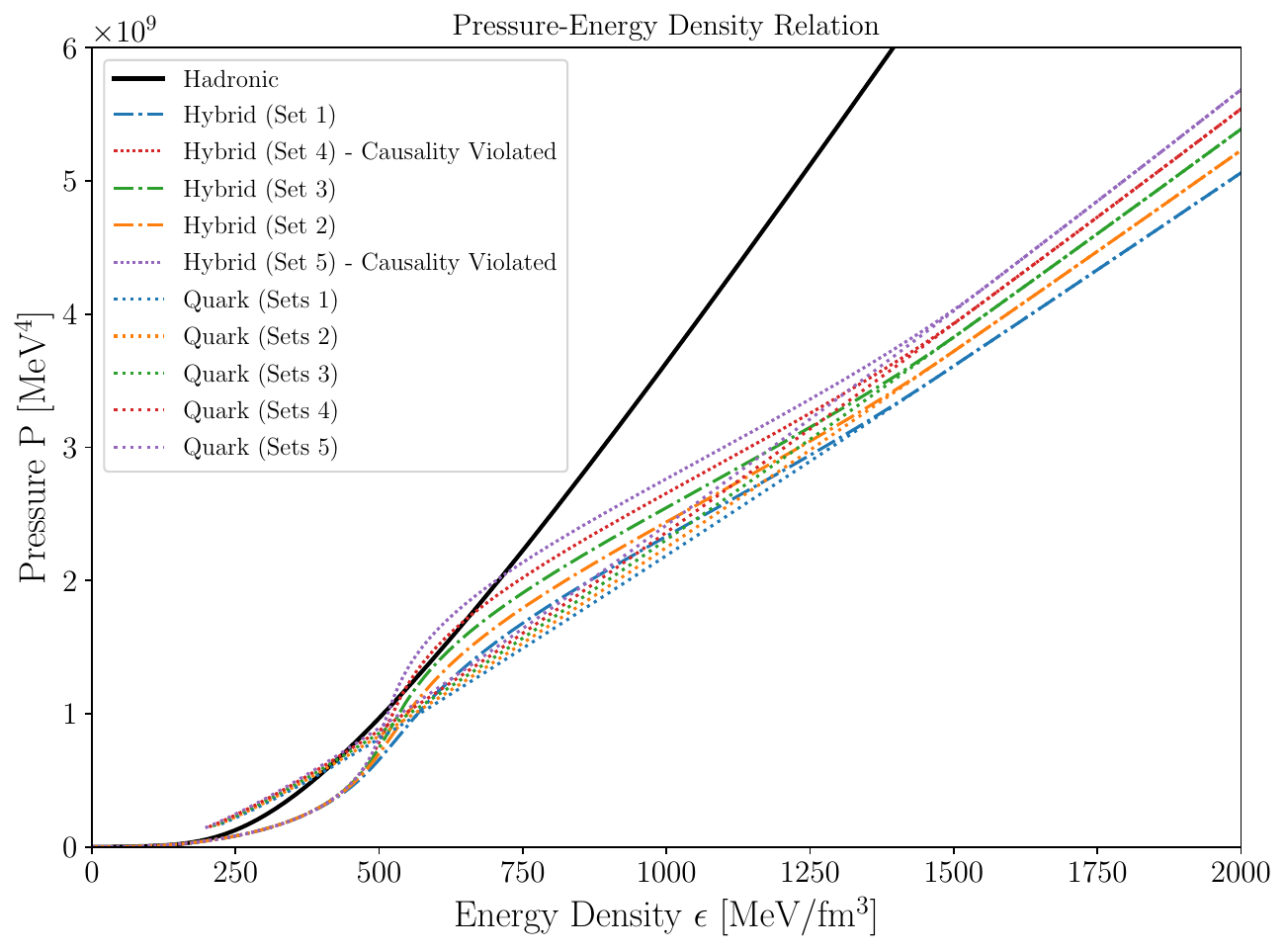}
        \centerline{(b) Pressure vs. Energy Density}
    \end{minipage}

    \vspace{1em}

    \begin{minipage}[b]{0.48\textwidth}
        \centering
        \includegraphics[width=\textwidth]{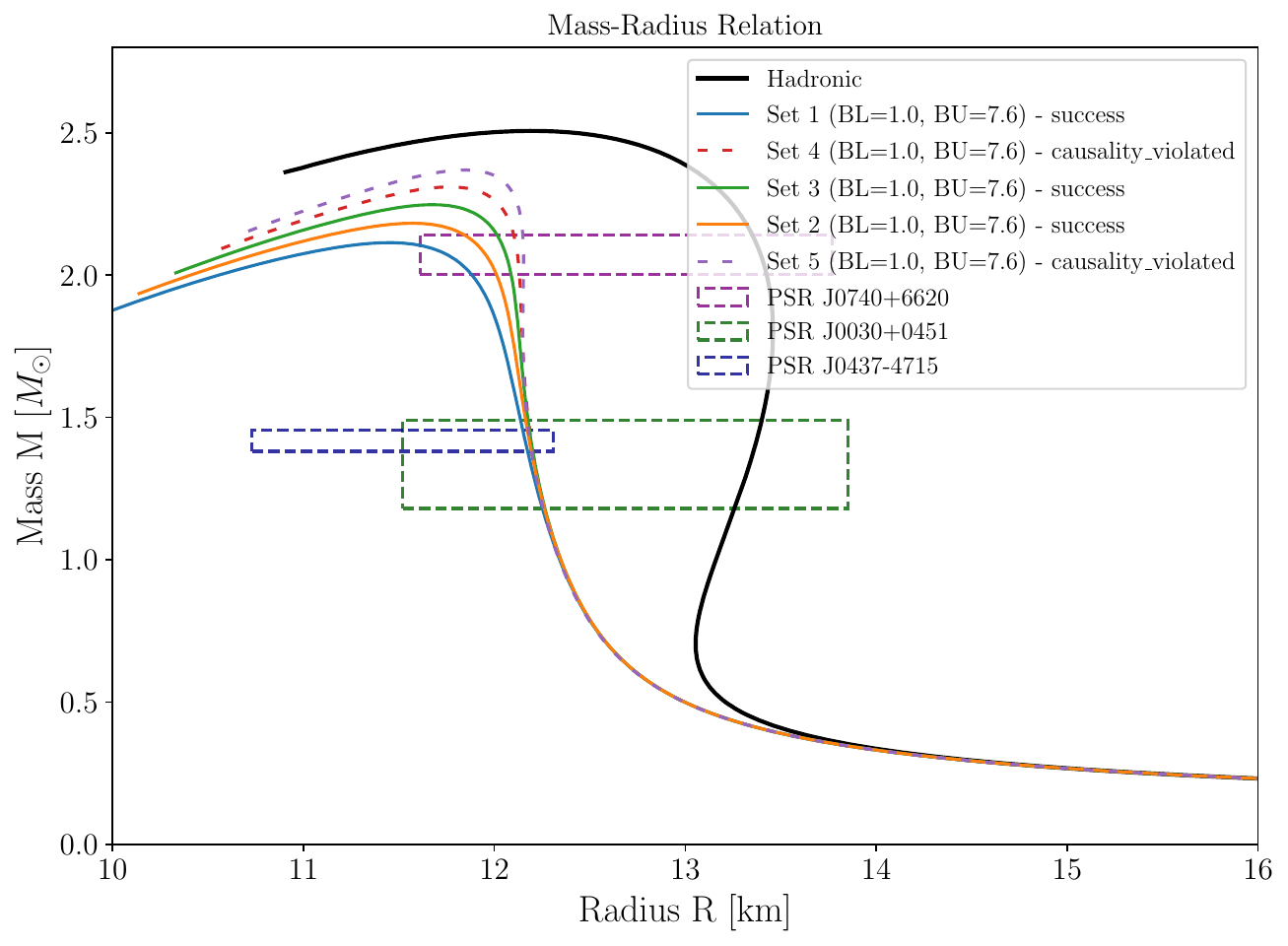}
        \centerline{(c) Mass-Radius Relation}
    \end{minipage}
    \hfill
    \begin{minipage}[b]{0.48\textwidth}
        \centering
        \includegraphics[width=\textwidth]{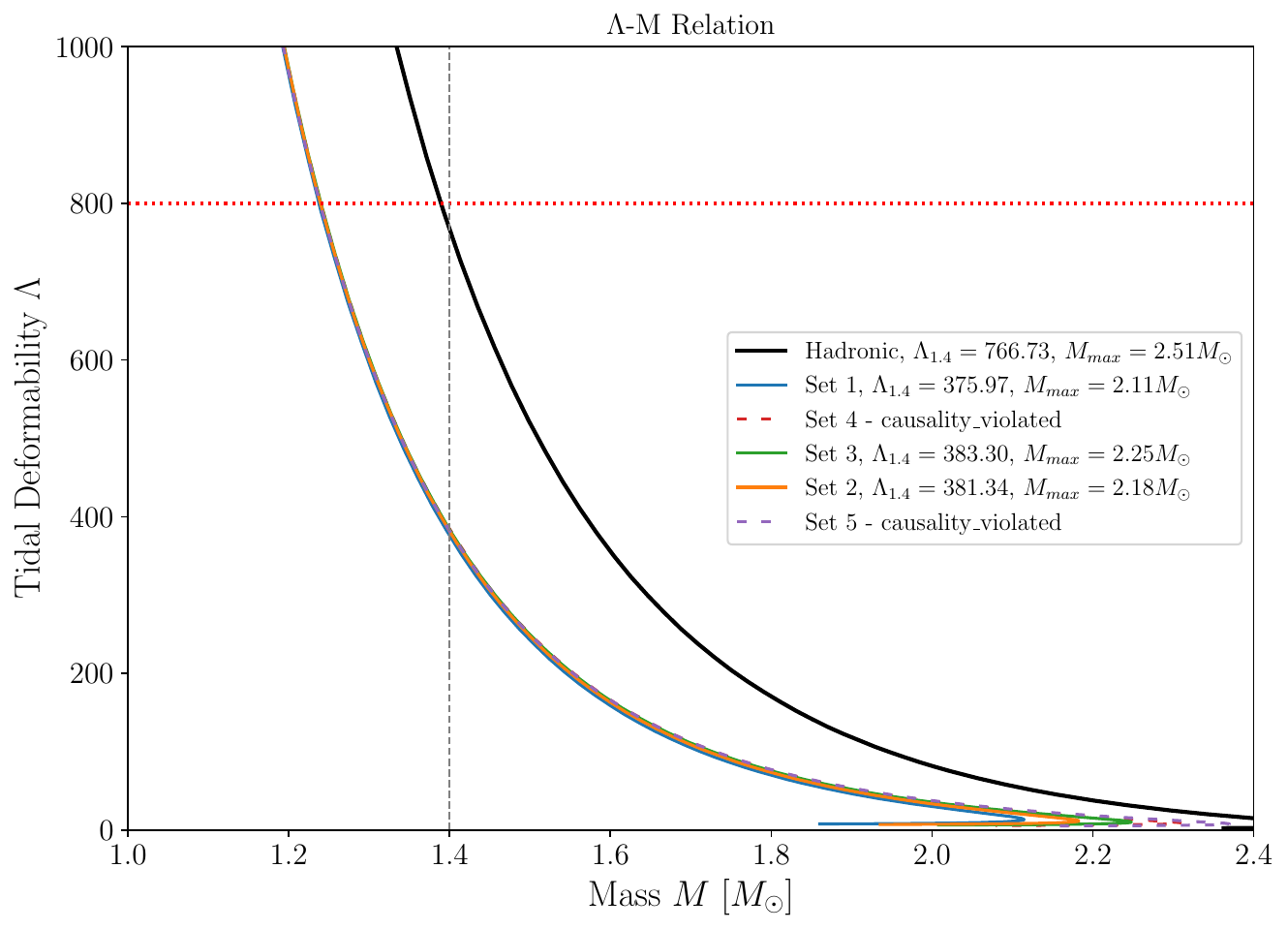}
        \centerline{(d) Tidal Deformability vs. Mass}
    \end{minipage}
    \caption{Impact of the vector coupling constant $G_V$ on the EOS and macroscopic properties of neutron stars, relative to the benchmark Set 3. The comparison is shown for five sets with $G_V/G_S$ ratios from 0.13 (Set 1) to 0.33 (Set 5). Panels (a) and (b) show how increasing $G_V$ stiffens the EOS at high densities and raises the sound speed peak. Panels (c) and (d) show how this translates to larger maximum masses while leaving $R_{1.4}$ and $\Lambda_{1.4}$ largely unchanged.}
    \label{fig:gv_comparison_combined_new}
    \end{figure*}

    To better visualize the impact of $G_V$, Figure \ref{fig:gv_comparison_combined_new} provides a comprehensive comparison of the five parameter sets at both the EOS and macroscopic levels.

    A detailed analysis of Figure \ref{fig:gv_comparison_combined_new} reveals the following:
    \begin{itemize}
        \item \textbf{Speed of Sound (Panel a)} clearly illustrates two effects of $G_V$. First, in the high-energy density (quark matter) region ($\epsilon \gtrsim 750 \text{ MeV/fm}^3$), the magnitude of $v_s^2/c^2$ is directly proportional to $G_V$. This demonstrates the role of vector repulsion in stiffening the matter at extreme densities. Second, the peak of the sound speed, which occurs in the crossover region (around $\epsilon \approx 500 \text{ MeV/fm}^3$), also rises sharply with $G_V$. This non-monotonic behavior and peak structure in the speed of sound are consistent with recent studies suggesting the plausible presence of a new state of matter in massive neutron stars \autocite{Han_2023_NewState}. For Sets 4 and 5, this peak significantly exceeds the causality limit $v_s^2/c^2 = 1$, confirming the findings from Table \ref{tab:gv_comparison_properties_new} that these EOSs are physically invalid.

        \item \textbf{Pressure-Energy Density Relation (Panel b)} corroborates this. While all curves overlap in the hadronic and crossover regions, they diverge at high densities. A larger $G_V$ (e.g., Set 5) results in a higher pressure at the same energy density, indicating a stiffer EOS. The quark matter curves (dotted lines) also clearly show this stiffening trend.

        \item \textbf{Mass-Radius Relation (Panel c)} shows how the EOS stiffness translates directly to the star's mass-bearing capacity. A stiffer EOS (larger $G_V$) supports a larger maximum mass. As $G_V/G_S$ increases from 0.13 to 0.33, $M_{max}$ grows from $2.11 M_\odot$ to $2.37 M_\odot$. The dashed lines (Sets 4 and 5) represent the acausal solutions. The physically valid curves (Sets 1, 2, 3) all successfully pass through the observational constraints from PSR J0740+6620, PSR J0030+0451, and PSR J0437-4715.

        \item \textbf{Tidal Deformability vs. Mass (Panel d)} confirms the analysis from the table. All five curves are clustered almost indistinguishably in the intermediate-mass range. At $1.4 M_\odot$, all predicted $\Lambda_{1.4}$ values are around $380$, far below the upper limit of $800$ from GW170817. This highlights that $M_{max}$ is highly sensitive to $G_V$, whereas $\Lambda_{1.4}$ and $R_{1.4}$ are not, for this parameter space.
    \end{itemize}

    In summary, the vector coupling constant $G_V$ is the key parameter for controlling the maximum mass of the hybrid star. This analysis demonstrates that we can use $G_V$ to ensure the model supports massive pulsars like PSR J0740+6620. However, this parameter is strongly constrained by causality, which provides a physical upper limit on its value (in this model, $G_V/G_S \lesssim 0.27$). Within the allowed, causal range, $G_V$ has minimal impact on the radius and tidal deformability of $1.4 M_\odot$ stars.

\section{Impact of the Phase Transition Endpoint $BU$ on Neutron Star Macroscopic Properties}
\label{sec:bu-impact}
    Besides the vector coupling constant $G_V$, which determines the "stiffness" of quark matter, the extent of the hadron-quark crossover region is also a key factor affecting the macroscopic properties of neutron stars. In this section, we again use our fiducial benchmark parameters ($G_S\Lambda^2=1.970$, $G_V/G_S=0.230$, $BL=1.0$) and vary only the phase transition endpoint coefficient $BU$. The parameter $BU$ defines the baryon number density at which matter fully transitions to the quark phase ($n_B = BU \times n_0$), so changing $BU$ is equivalent to altering the width of the phase transition window.

\subsection{Parameter Sets and Comparison of Key Properties}
\label{subsec:bu-params-table}
    We select five representative values for $BU$: 5.6, 6.6, 7.6 (our fiducial benchmark), 8.6, and 9.6. All parameter sets use the same phase transition starting point, $BL=1.0$. Table \ref{tab:bu_comparison_properties_new} quantitatively summarizes the key macroscopic properties for these parameter sets, based on the data in our summary file.

\begin{table}[htbp]
    \centering
    \caption{Comparison of Key Neutron Star Macroscopic Properties for Different $BU$ Parameters. All sets share the same NJL parameters ($G_S\Lambda^2=1.970$, $G_V/G_S=0.230$) and $BL=1.0$.
    \newline
    \textbf{Note:} These parameter sets exhibit a causality violation ($v_s^2/c^2 > 1$) in the crossover region (see Figure \ref{fig:bu_comparison_combined_new}a); thus, their results are for theoretical trend reference only and are not physically acceptable.
    }
    \label{tab:bu_comparison_properties_new}
    \small
    \begin{tabular}{lccccc}
        \hline\hline
        \textbf{Parameter Set} & \textbf{$BU$ Coefficient} & \textbf{$M_{max}$ ($M_\odot$)} & \textbf{$R_{1.4}$ (km)} & \textbf{$\Lambda_{1.4}$} & \textbf{Status} \\
        \hline
        Set 1 & 5.6 & 2.24 & 12.47 & 459.2 & Success \\
        Set 2 & 6.6 & 2.25 & 12.33 & 417.4 & Success \\
        Set 3 & 7.6 & 2.25 & 12.20 & 383.3 & Success \\
        Set 4$^{a}$ & 8.6 & 2.24 & 12.08 & 355.5 & Violated \\
        Set 5$^{a}$ & 9.6 & 2.22 & 11.98 & 332.1 & Violated \\
        \hline\hline
    \end{tabular}
\end{table}

    The table reveals a set of physical trends that are distinctly different from those of $G_V$.
    First, in the physically causal region (Sets 1-3), the maximum mass $M_{max}$ is almost entirely insensitive to $BU$, remaining constant at $2.24-2.25 M_\odot$.
    
    Second, $BU$ is the primary parameter for tuning the properties of $1.4 M_\odot$ stars. There is a strong \textbf{negative correlation}: as $BU$ increases (i.e., the transition region widens), the $R_{1.4}$ \textbf{decreases} (from 12.47 km to 12.20 km) and $\Lambda_{1.4}$ \textbf{decreases} significantly (from 459 to 383). This highlights that a wider crossover window produces a softer EOS in the intermediate density range, leading to more compact stars.
    
    Third, similar to $G_V$, $BU$ is also constrained by causality. As $BU$ increases, the crossover EOS eventually becomes acausal. Sets 4 and 5 ($BU=8.6, 9.6$) are rendered physically invalid by causality violations, establishing a physical upper bound on the width of the transition region ($BU \lesssim 7.6$).

\subsection{Graphical Analysis of the EOS and Macroscopic Properties}
\label{subsec:bu-plots-analysis}
\begin{figure*}[htbp]
    \centering
    \begin{minipage}[b]{0.48\textwidth}
        \centering
        \includegraphics[width=\textwidth]{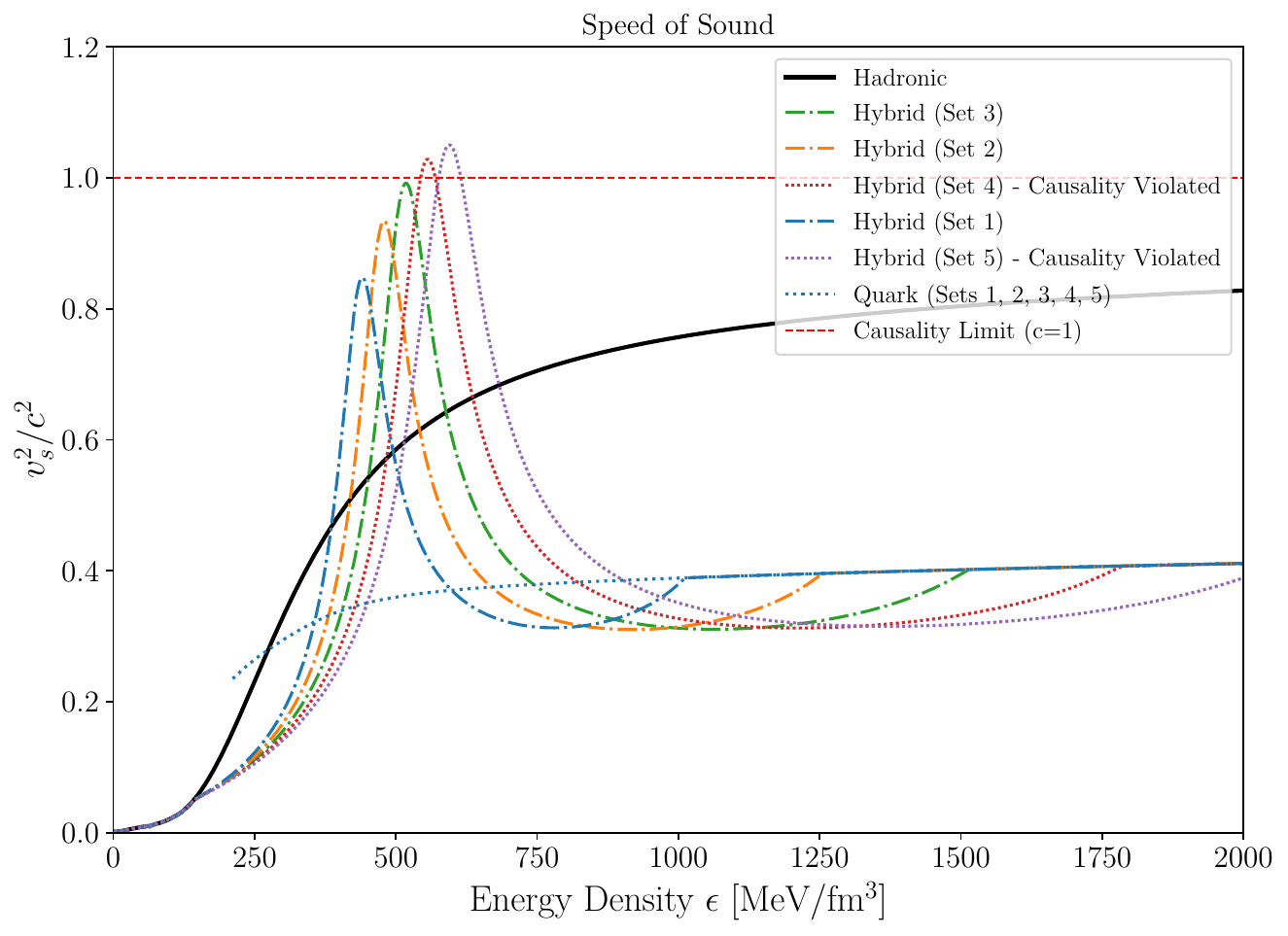}
        \centerline{(a) Speed of Sound vs. Energy Density}
    \end{minipage}
    \hfill
    \begin{minipage}[b]{0.48\textwidth}
        \centering
        \includegraphics[width=\textwidth]{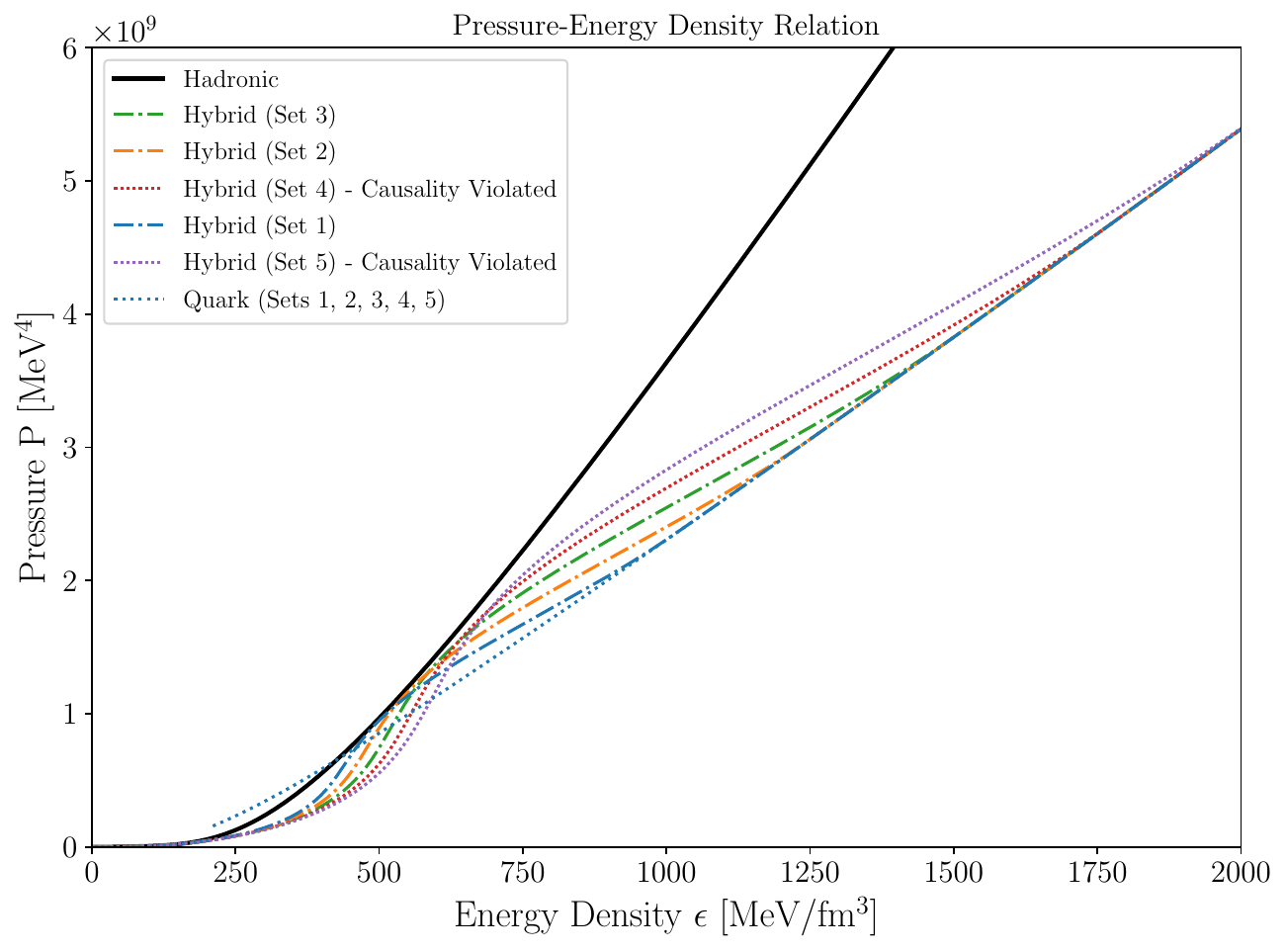}
        \centerline{(b) Pressure vs. Energy Density}
    \end{minipage}

    \vspace{1em}

    \begin{minipage}[b]{0.48\textwidth}
        \centering
        \includegraphics[width=\textwidth]{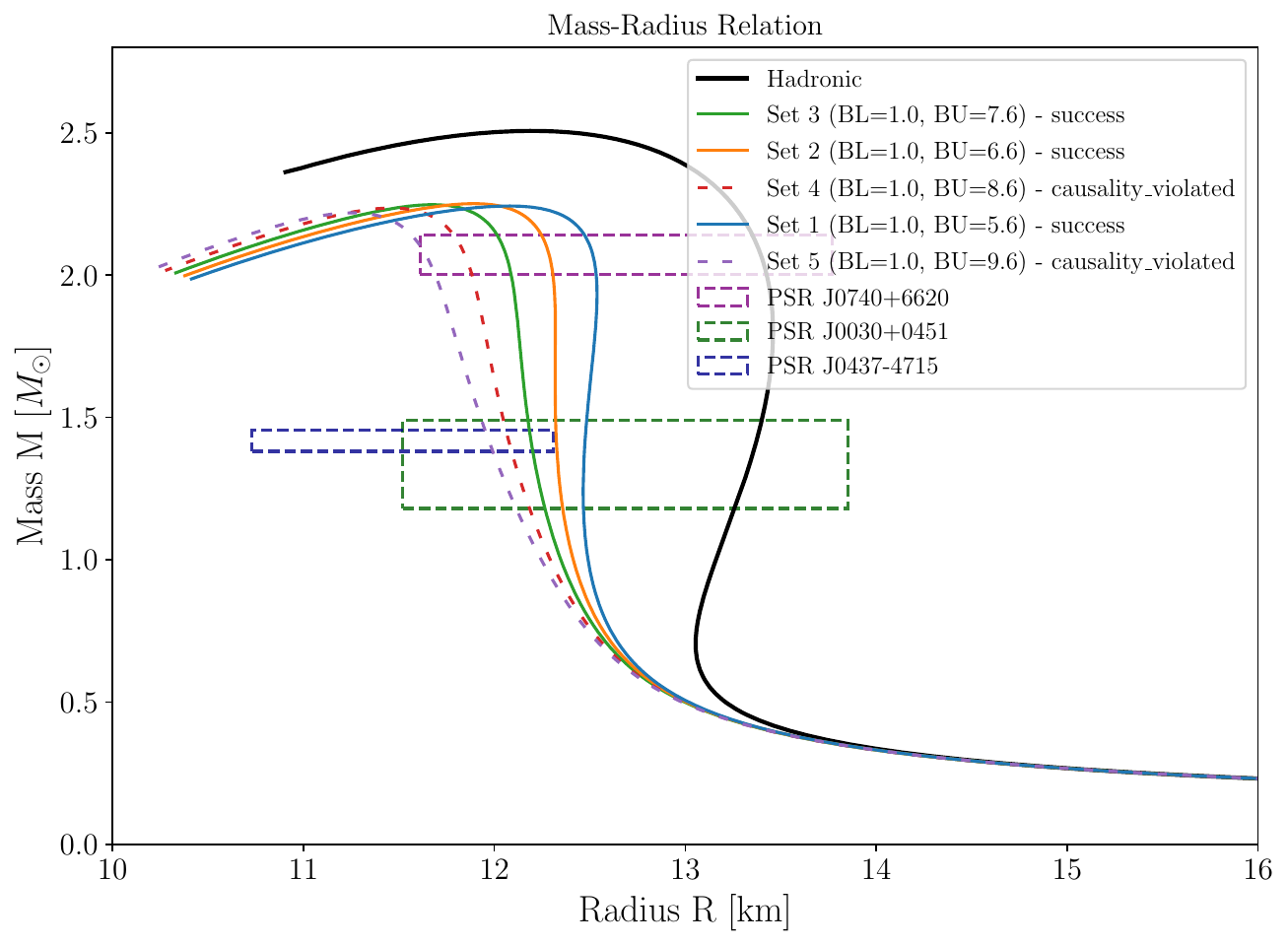}
        \centerline{(c) Mass-Radius Relation}
    \end{minipage}
    \hfill
    \begin{minipage}[b]{0.48\textwidth}
        \centering
        \includegraphics[width=\textwidth]{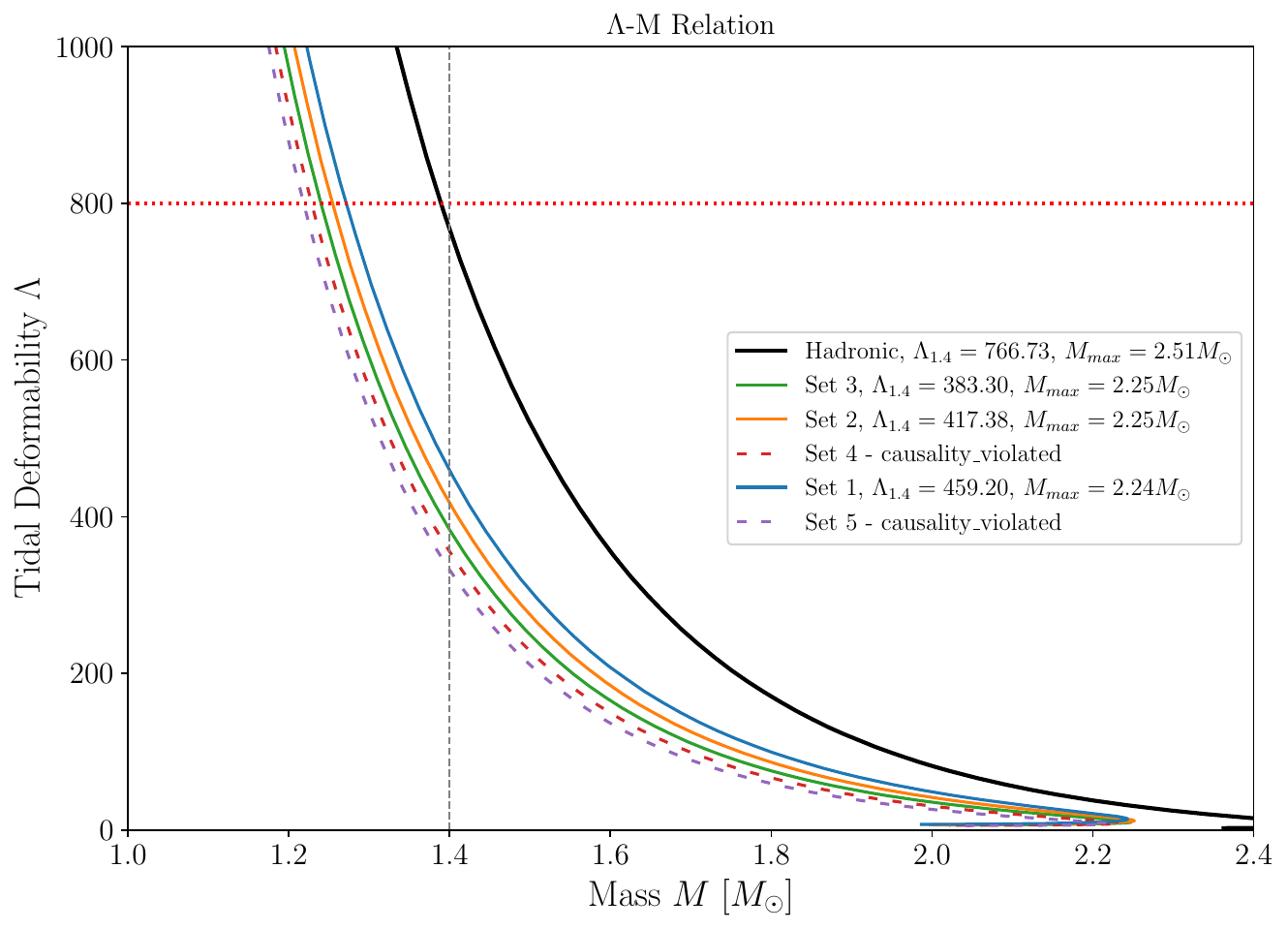}
        \centerline{(d) Tidal Deformability vs. Mass}
    \end{minipage}
    \caption{Impact of the phase transition endpoint $BU$ on the EOS and macroscopic properties of neutron stars. All sets share the same NJL parameters and $BL=1.0$. The comparison is shown for five sets with $BU$ coefficients from 5.6 (Set 1) to 9.6 (Set 5). Panels (a) and (b) show how increasing $BU$ (widening the crossover) makes the EOS softer and raises the sound speed peak. Panels (c) and (d) show how this softening translates to smaller radii and tidal deformabilities.}
    \label{fig:bu_comparison_combined_new}
    \end{figure*}

    To better visualize the impact of $BU$, Figure \ref{fig:bu_comparison_combined_new} provides a comprehensive comparison of the five parameter sets at both the EOS and macroscopic levels .

    From Figure \ref{fig:bu_comparison_combined_new}, we can clearly uncover the mechanism by which $BU$ affects the EOS:
    \begin{itemize}
        \item \textbf{Pressure-Energy Density Relation (Panel b)} clearly shows that all quark EOSs (dotted line) are identical, as are the hadronic EOSs (solid black line). As $BU$ increases, the interpolation window widens, forcing the hybrid EOS to "sag" or become \textbf{softer} in the intermediate density range before rejoining the stiff hadronic curve. Set 1 (BU=5.6) is the stiffest, while Set 3 (BU=7.6) is softer, and the causality-violating Sets 4 and 5 are softer still.

        \item \textbf{Mass-Radius Relation (Panel c)} directly reflects this softness. The stiffest EOS (Set 1, BU=5.6) produces the largest radius ($R_{1.4} \approx 12.47$ km). As $BU$ increases, the EOS becomes softer, and the M-R curve shifts to the left, producing a smaller radius. The benchmark Set 3 (BU=7.6) is the most compact of the causal sets ($R_{1.4} \approx 12.20$ km). The maximum mass $M_{max}$ is only slightly affected within the causal region.
        
        \item \textbf{Tidal Deformability vs. Mass (Panel d)} confirms this trend perfectly. A larger $BU$ (softer intermediate EOS) leads to a smaller, less deformable star. $\Lambda_{1.4}$ systematically \textbf{decreases} as $BU$ increases, from 459.2 (Set 1) down to 383.3 (Set 3).

        \item \textbf{Speed of Sound (Panel a)} reveals the physical constraint. The peak of the sound speed, which occurs in the crossover region, becomes progressively \textit{higher and sharper} as $BU$ increases (i.e., as the interpolation window widens). For $BU=8.6$ (Set 4) and $BU=9.6$ (Set 5), this peak clearly violates the causality limit $v_s^2/c^2 = 1$. This establishes a physical \textit{upper bound} on the width of the crossover region ($BU \lesssim 7.6$) for this model.
    \end{itemize}

    In summary, the phase transition endpoint parameter $BU$ is the most effective tool for tuning the radius and tidal deformability of $1.4 M_\odot$ neutron stars. It primarily adjusts the EOS softness in the intermediate-density range without significantly affecting the maximum mass. The observation that a wider transition (larger $BU$) produces a *softer* EOS, resulting in a *smaller* radius and *smaller* tidal deformability, is a key finding. This parameter is strongly constrained by causality, which provides an upper limit on how wide the transition can be.

\section{Impact of the Scalar Coupling Factor $G_S\Lambda^2$ on Neutron Star Macroscopic Properties}
\label{sec:glambda2-impact}
    Finally, we investigate the impact of the scalar coupling strength, parameterized by the dimensionless factor $G_S\Lambda^2$. This parameter directly controls the strength of the scalar interaction $G_S$ and thus influences the generation of dynamical quark mass. As with the previous sections, we use our fiducial benchmark parameters ($G_V/G_S=0.230$, $BL=1.0$, $BU=7.60$) and vary only $G_S\Lambda^2$.

\subsection{Parameter Sets and Comparison of Key Properties}
\label{subsec:glambda2-params-table}
    We select five representative values for $G_S\Lambda^2$: 1.770, 1.870, 1.970 (our fiducial benchmark, Set 3), 2.070, and 2.170. All other parameters are held fixed. Table \ref{tab:glambda2_comparison_properties_new} summarizes the key macroscopic properties for these parameter sets, based on our data.It is important to note that in standard NJL model phenomenology, the parameters $G_S$, $\Lambda$, and $m_{u, d}$ are typically fitted simultaneously to reproduce physical vacuum observables, such as the pion mass $m_\pi$ and decay constant $f_\pi$. In this study, we adopt \textbf{Set 3} ($G_S\Lambda^2=1.970$) as the physical baseline, which is strictly calibrated to these empirical values. For the other sets (Sets 1, 2, 4, and 5), we intentionally vary the dimensionless coupling $G_S\Lambda^2$ while keeping the cutoff $\Lambda$ and current quark mass $m_{u, d}$ fixed. Although this approach implies that the vacuum properties for the variant sets deviate slightly from the experimental data, it allows us to \textit{isolate} the specific impact of the scalar interaction strength on the high-density EOS and the stiffness of the quark matter, decoupling it from the effects of varying the momentum cutoff scale. Therefore, Sets 1, 2, 4, and 5 should be interpreted as a sensitivity analysis of the model's stiffness to the effective interaction strength, centered around the physical benchmark Set 3.

\begin{table}[htbp]
    \centering
    \caption{Comparison of Key Neutron Star Macroscopic Properties for Different $G_S\Lambda^2$ Parameters. All sets share the same $G_V/G_S=0.230$, $BL=1.0$, and $BU=7.60$.
    \textbf{Note (a):} These parameter sets exhibit a causality violation ($v_s^2/c^2 > 1$), as shown in Figure \ref{fig:glambda2_comparison_combined_new}a; thus, their results are for theoretical trend reference only and are not physically acceptable.
    }
    \label{tab:glambda2_comparison_properties_new}
    \small
    \begin{tabular}{lccccc}
        \hline\hline
        \textbf{Parameter Set} & \textbf{$G_S\Lambda^2$} & \textbf{$M_{max}$ ($M_\odot$)} & \textbf{$R_{1.4}$ (km)} & \textbf{$\Lambda_{1.4}$} & \textbf{Status} \\
        \hline
        Set 1 & 1.770 & 2.09 & 11.88 & 310.6 & Success  \\
        Set 2 & 1.870 & 2.15 & 12.02 & 341.6 & Success  \\
        Set 3 & 1.970 & 2.25 & 12.20 & 383.3 & Success  \\
        Set 4$^{a}$ & 2.070 & 2.39 & 12.39 & 434.9 & Violated  \\
        Set 5$^{a}$ & 2.170 & 2.60 & 12.59 & 495.1 & Violated  \\
        \hline\hline
    \end{tabular}
\end{table}

    The table reveals a complex set of correlations. First, $G_S\Lambda^2$ has a strong, positive impact on $M_{max}$. As the scalar coupling increases from 1.770 to 2.170, $M_{max}$ systematically grows from $2.09 M_\odot$ to $2.60 M_\odot$. This indicates that a stronger scalar interaction leads to a stiffer high-density EOS.
    
    Second, unlike $G_V$, $G_S\Lambda^2$ significantly impacts the properties of $1.4 M_\odot$ stars, but in a divergent way. As $G_S\Lambda^2$ increases, the radius $R_{1.4}$ \textbf{increases} (from 11.88 km to 12.59 km), and the tidal deformability $\Lambda_{1.4}$ also \textbf{increases} (from 310.6 to 495.1). This demonstrates that, for this parameter, a stiffer EOS (larger $G_S\Lambda^2$) results in both a larger radius and a larger tidal deformability at $1.4 M_\odot$.
    
    Third, $G_S\Lambda^2$ is also constrained by causality. As seen in Table \ref{tab:glambda2_comparison_properties_new} and the figures, values of $G_S\Lambda^2 \ge 2.070$ (Sets 4 and 5) cause the sound speed to exceed $c$, rendering them physically invalid and establishing an upper bound for this parameter.

\subsection{Graphical Analysis of the EOS and Macroscopic Properties}
\label{subsec:glambda2-plots-analysis}
\begin{figure*}[htbp]
    \centering
    \begin{minipage}[b]{0.48\textwidth}
        \centering
        \includegraphics[width=\textwidth]{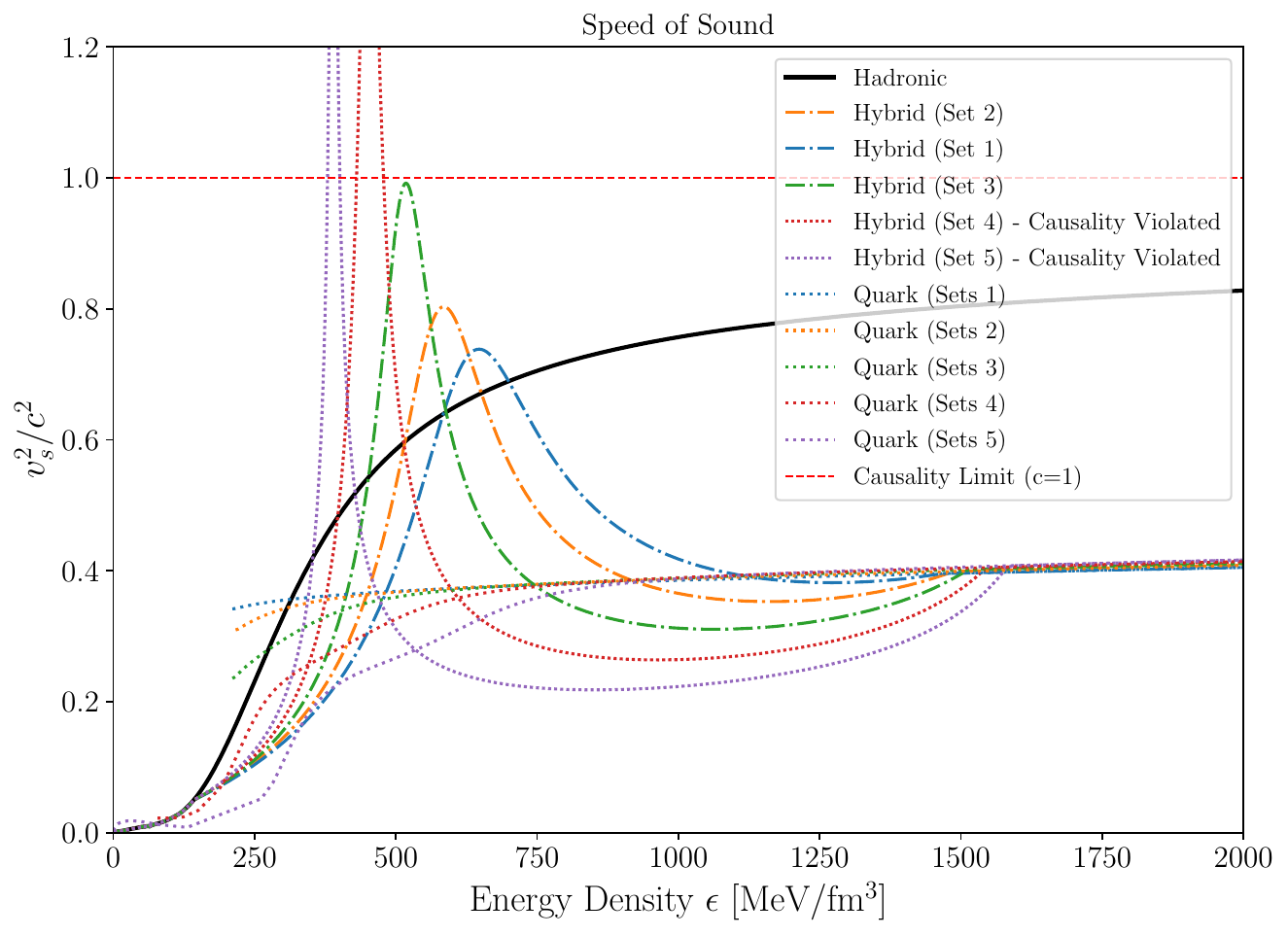}
        \centerline{(a) Speed of Sound vs. Energy Density}
    \end{minipage}
    \hfill
    \begin{minipage}[b]{0.48\textwidth}
        \centering
        \includegraphics[width=\textwidth]{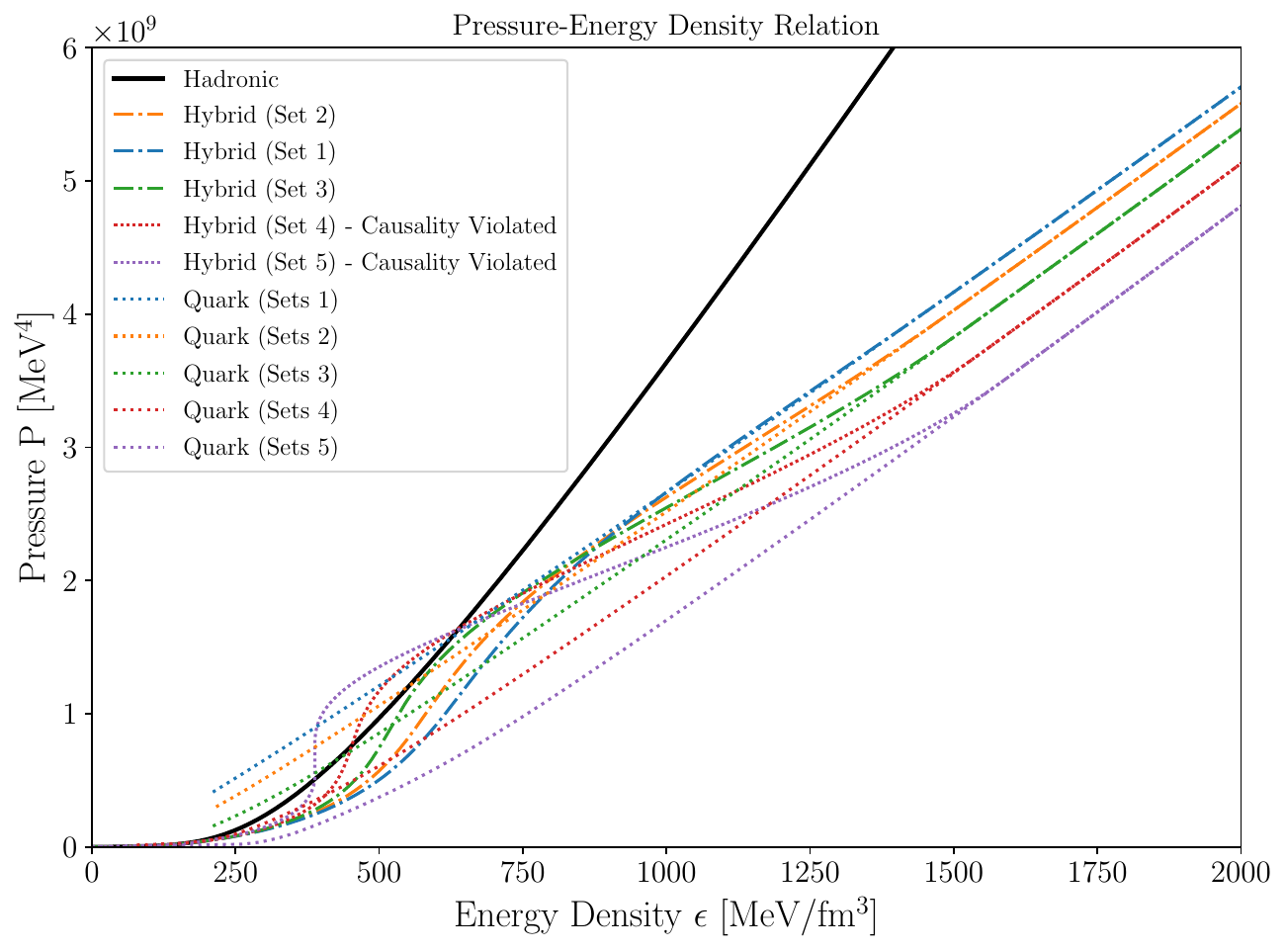}
        \centerline{(b) Pressure vs. Energy Density}
    \end{minipage}

    \vspace{1em}

    \begin{minipage}[b]{0.48\textwidth}
        \centering
        \includegraphics[width=\textwidth]{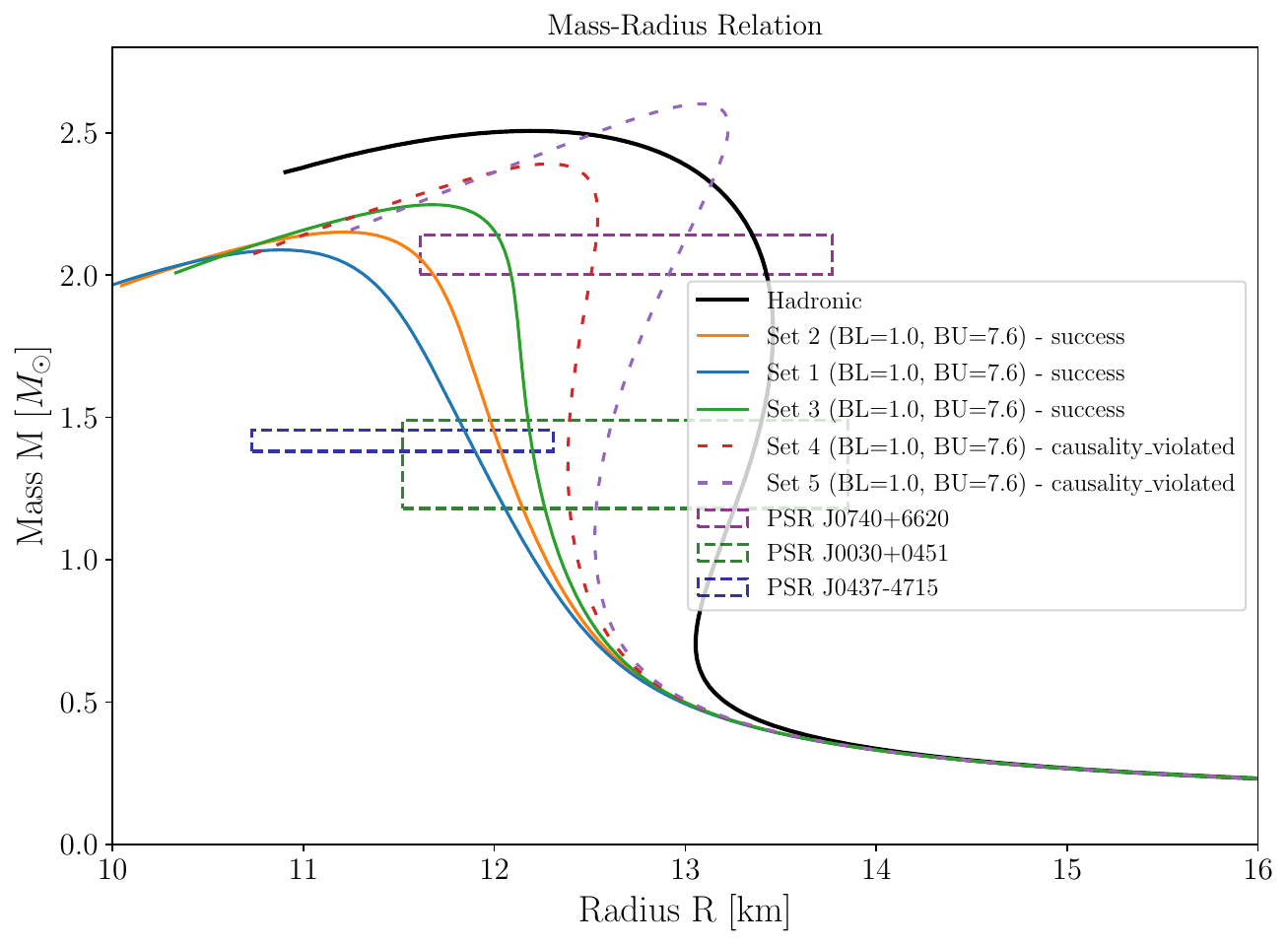}
        \centerline{(c) Mass-Radius Relation}
    \end{minipage}
    \hfill
    \begin{minipage}[b]{0.48\textwidth}
        \centering
        \includegraphics[width=\textwidth]{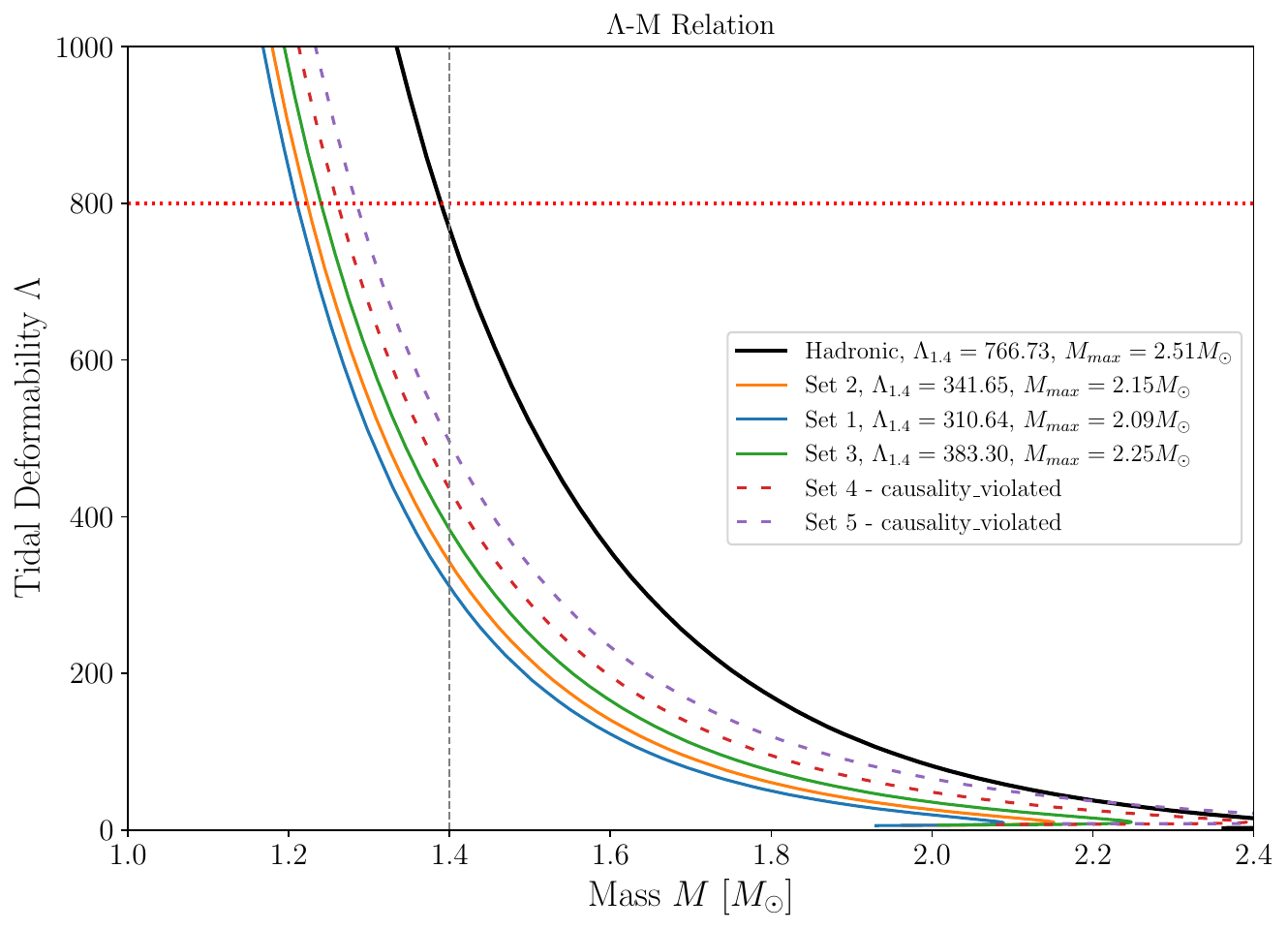}
        \centerline{(d) Tidal Deformability vs. Mass}
    \end{minipage}
    \caption{Impact of the scalar coupling factor $G_S\Lambda^2$ on the EOS and macroscopic properties of neutron stars. All sets share the same $G_V/G_S$, $BL$, and $BU$,. Panels (a) and (b) show how increasing $G_S\Lambda^2$ stiffens the quark matter EOS and raises the sound speed peak. Panels (c) and (d) show the resulting increase in $M_{max}$, $R_{1.4}$, and $\Lambda_{1.4}$.}
    \label{fig:glambda2_comparison_combined_new}
    \end{figure*}

    Figure \ref{fig:glambda2_comparison_combined_new} provides a graphical analysis of these five parameter sets, illustrating the underlying EOS behavior.

    A detailed analysis of Figure \ref{fig:glambda2_comparison_combined_new} reveals the following mechanisms:
    \begin{itemize}
        \item \textbf{Pressure-Energy Density Relation (Panel b)} shows that a larger $G_S\Lambda^2$ (stronger scalar coupling) leads to a \textit{stiffer} quark matter EOS. This is evident from the quark phase curves (dotted lines), where Set 5 (highest $G_S\Lambda^2$) has a higher pressure at a given energy density than Set 1 (lowest $G_S\Lambda^2$). The hybrid EOS curves (dashed/solid lines) inherit this stiffness at high densities.

        \item \textbf{Speed of Sound (Panel a)} illustrates two simultaneous effects. First, consistent with the $P-\epsilon$ relation, the sound speed in the pure quark phase (high density) is higher for a larger $G_S\Lambda^2$. Second, the sound speed peak in the crossover region also increases dramatically with $G_S\Lambda^2$, leading to the causality violations for Set 4 and Set 5.

        \item \textbf{Mass-Radius Relation (Panel c)} shows the macroscopic consequences. The stiffer high-density EOS from a larger $G_S\Lambda^2$ results in a higher $M_{max}$. In the $1.4 M_\odot$ region, the trend is consistent: a larger $G_S\Lambda^2$ (e.g., Set 3) produces a *larger* radius ($R_{1.4} \approx 12.20$ km) than a smaller $G_S\Lambda^2$ (e.g., Set 1, $R_{1.4} \approx 11.88$ km).

        \item \textbf{Tidal Deformability vs. Mass (Panel d)} further highlights this behavior. $\Lambda_{1.4}$ \textit{increases} with $G_S\Lambda^2$, rising from 310.6 (Set 1) to 383.3 (Set 3). This shows that, for this parameter, a stiffer EOS leads to both a larger radius and a larger tidal deformability, in line with typical EOS behavior.
    \end{itemize}

    In summary, the scalar coupling $G_S\Lambda^2$ plays a critical role, stiffening the EOS at nearly all densities. This simultaneously increases $M_{max}$, $R_{1.4}$, and $\Lambda_{1.4}$. Like the other parameters, it is strongly constrained by causality, which provides an upper limit ($G_S\Lambda^2_{\text{actor}} \lesssim 1.970$) in this model.

\section{Conclusion}
\label{sec:conclusion}
In this paper, we have constructed and systematically studied an Equation of State (EOS) for hadron-quark hybrid stars. The model describes hadronic matter using the unified DDME2 density-dependent relativistic mean-field model and high-density quark matter using a two-flavor Nambu-Jona-Lasinio (NJL) model. A core element of our work was the use of a quintic polynomial interpolation method ensuring $C^2$ continuity to build a smooth hadron-quark crossover. Our primary goal was to test whether this hybrid star model could resolve the central tension in multi-messenger astronomy: the need for a "stiff" EOS at high densities to support massive pulsars, and a "soft" EOS at intermediate densities to satisfy radius and tidal deformability constraints.

We have reached the following core conclusions:
\begin{enumerate}
    \item Through a systematic parameter investigation, we established a \textbf{fiducial benchmark parameter set} (Set 3: $G_S\Lambda^2=1.970$, $G_V/G_S=0.23$, $BU=7.60$, $BL=1.0$) that successfully satisfies all current key observational constraints.
    
    \item Our fiducial benchmark model predicts a maximum mass of $M_{\text{max}} \approx 2.25 M_{\odot}$. This result is not only well above the limit set by PSR J0740+6620 but also aligns remarkably with the precise inference of $M_{TOV} = 2.25^{+0.08}_{-0.07} M_{\odot}$ derived from multi-messenger data \autocite{Fan_2024_MaxMass}. Concurrently, it produces a compact radius ($R_{1.4} \approx 12.20$ km) and low tidal deformability ($\Lambda_{1.4} \approx 383$), in excellent agreement with NICER and GW170817 data. The speed of sound in our model exhibits a non-monotonic peak, consistent with the characteristics of a new state of matter \autocite{Han_2023_NewState}.
    
    \item A key finding of this work is the \textbf{necessity of a low-density crossover onset}. As discussed in Section \ref{subsec:njl-parameters}, a transition onset at $BL \approx 1.0$ is required to satisfy the compact radius constraints of pulsars like PSR J0437-4715. Models with a delayed transition ($BL \ge 1.2$) fail to reconcile the tension between high mass and compact radius. This implies that the \textbf{pure hadronic description based on point particles likely reaches its validity limit near saturation density}, allowing for the gradual percolation of quark degrees of freedom to accommodate modern multi-messenger constraints.
    
    \item This low-density transition works in tandem with the model's other key parameters, which we found have distinct and complementary roles (Sections \ref{sec:gv-impact}-\ref{sec:glambda2-impact}). The vector coupling $G_V$ (Section \ref{sec:gv-impact}) is the primary parameter for tuning $M_{\text{max}}$ but has minimal impact on $R_{1.4}$. Conversely, the crossover endpoint $BU$ (Section \ref{sec:bu-impact}) is the main tool for tuning $R_{1.4}$, where a wider crossover (larger $BU$) leads to a *softer* EOS and thus a *smaller* radius. Finally, the scalar coupling $G_S\Lambda^2$ (Section \ref{sec:glambda2-impact}) acts as a global stiffening parameter.
    
    \item We found that all three parameters ($G_V$, $BU$, $G_S\Lambda^2$) are strongly constrained by \textbf{causality}, which provides a physical upper bound on their values. This combination of constraints and distinct control mechanisms is the core strength of the RMF-NJL crossover model.
\end{enumerate}

In summary, this work validates the RMF-NJL crossover model as a self-consistent and robust framework. We have shown that by establishing a necessary low-density transition and leveraging the distinct, causality-constrained roles of the model's key parameters, this approach successfully reconciles the full suite of multi-messenger observational data.

\appendix
\section{Solving for the Interpolation Coefficients of the Hybrid EOS}
\label{appendix:interpolation_coeffs}

In Section \ref{sec:eos-construction} of this article, a quintic polynomial is employed to construct a smooth crossover region from the hadronic phase to the quark phase. This polynomial describes the relationship between pressure $P$ and baryon chemical potential $\mu_B$ within the transition interval [$\mu_{BL}, \mu_{BU}$]:
\begin{equation}
    P(\mu_B) = \sum_{m=0}^{5} C_{m} \mu_{B}^{m} = C_0 + C_1 \mu_B + C_2 \mu_B^2 + C_3 \mu_B^3 + C_4 \mu_B^4 + C_5 \mu_B^5
\end{equation}
To ensure that the entire hybrid equation of state is $C^2$ continuous—meaning that the pressure, baryon number density ($\rho_B = dP/d\mu_B$), and its first derivative ($d\rho_B/d\mu_B = d^2P/d\mu_B^2$) are all continuous at the boundaries—we impose six boundary conditions.

These six boundary conditions form a system of linear equations for the six unknown coefficients ($C_0, \dots, C_5$), specified as follows:
\begin{enumerate}
    \item $P(\mu_{BL}) = P_H(\mu_{BL})$
    \item $P'(\mu_{BL}) = \rho_{B,H}(\mu_{BL})$
    \item $P''(\mu_{BL}) = \chi_{B,H}(\mu_{BL})$
    \item $P(\mu_{BU}) = P_Q(\mu_{BU})$
    \item $P'(\mu_{BU}) = \rho_{B,Q}(\mu_{BU})$
    \item $P''(\mu_{BU}) = \chi_{B,Q}(\mu_{BU})$
\end{enumerate}
Here, the subscripts $H$ and $Q$ denote quantities from the hadronic and quark phases, respectively, and $\chi_B = d\rho_B/d\mu_B$ is the baryon number susceptibility.

This system of equations can be written in matrix form as $\mathbf{A}\mathbf{x} = \mathbf{b}$, where $\mathbf{x} = [C_0, C_1, C_2, C_3, C_4, C_5]^T$ is the vector of coefficients to be solved for. The matrix $\mathbf{A}$ and vector $\mathbf{b}$ are given by:
\begin{equation}
\mathbf{A} = 
\begin{pmatrix}
1 & \mu_{BL} & \mu_{BL}^2 & \mu_{BL}^3 & \mu_{BL}^4 & \mu_{BL}^5 \\
0 & 1 & 2\mu_{BL} & 3\mu_{BL}^2 & 4\mu_{BL}^3 & 5\mu_{BL}^4 \\
0 & 0 & 2 & 6\mu_{BL} & 12\mu_{BL}^2 & 20\mu_{BL}^3 \\
1 & \mu_{BU} & \mu_{BU}^2 & \mu_{BU}^3 & \mu_{BU}^4 & \mu_{BU}^5 \\
0 & 1 & 2\mu_{BU} & 3\mu_{BU}^2 & 4\mu_{BU}^3 & 5\mu_{BU}^4 \\
0 & 0 & 2 & 6\mu_{BU} & 12\mu_{BU}^2 & 20\mu_{BU}^3
\end{pmatrix}
\end{equation}
\begin{equation}
\mathbf{b} = 
\begin{pmatrix}
P_H(\mu_{BL}) \\ \rho_{B,H}(\mu_{BL}) \\ \chi_{B,H}(\mu_{BL}) \\ P_Q(\mu_{BU}) \\ \rho_{B,Q}(\mu_{BU}) \\ \chi_{B,Q}(\mu_{BU})
\end{pmatrix}
\end{equation}
By solving this linear system, the polynomial coefficients $\mathbf{x} = \mathbf{A}^{-1}\mathbf{b}$ can be uniquely determined, thereby constructing a thermodynamically consistent and smooth crossover equation of state.

The system of linear equations $\mathbf{A}\mathbf{x} = \mathbf{b}$ is then solved numerically for the specific parameter sets chosen in this study (e.g., the fiducial benchmark Set 3).This procedure uniquely determines the coefficients $C_0, \dots, C_5$, thereby constructing a thermodynamically consistent and smooth crossover equation of state that satisfies all boundary conditions at $\mu_{BL}$ and $\mu_{BU}$.The resulting hybrid EOS ensures the continuity of pressure, baryon density, and susceptibility, which is crucial for the stability and physical validity of the neutron star structure calculations.

\section{Explicit Forms of the Tidal Perturbation Functions}
\label{appendix:tidal_functions}

In Section \ref{subsec:tidal-deformability} of this article, the master function $H(r)$, which describes the quadrupole deformation of a neutron star under an external tidal field, is obtained by solving a second-order ordinary differential equation (Eq. \ref{eq:H_equation}). This equation arises from perturbing the background spacetime of a spherically symmetric star within the framework of general relativity.

Specifically, this result is derived by linearizing the Einstein field equations under the Regge-Wheeler gauge, considering static, even-parity, quadrupole ($l=2$) perturbations. The coefficients $F(r)$ and $Q(r)$ in the equation are entirely determined by the unperturbed background spacetime (the solution to the TOV equations) and the equation of state (EOS) of the matter. Their explicit expressions are as follows \autocite{Hinderer_2008_TidalLoveNumbers, Hinderer_2010_TidalDeformabilityRealisticEOS}:

First, the background spacetime is described by the spherically symmetric Schwarzschild metric, which takes the form:
\begin{equation}
    ds^2 = -e^{2\nu(r)}dt^2 + e^{2\lambda(r)}dr^2 + r^2(d\theta^2 + \sin^2\theta d\phi^2)
\end{equation}
where the metric function $e^{2\lambda(r)} = (1-2GM(r)/r)^{-1}$. The terms $M(r)$ and $P(r)$ are the radial mass and pressure profiles, respectively, obtained by integrating the TOV equations (Eqs. \ref{eq:tov_pressure} and \ref{eq:tov_mass}). The derivative of the other metric function, $\nu(r)$, is given by:
\begin{equation}
    \nu'(r) = \frac{dP/dr}{\epsilon(r) + P(r)} = \frac{G(M(r) + 4\pi r^3 P(r))}{r(r - 2GM(r))}
\end{equation}
Here, $\epsilon(r)$ is the energy density corresponding to $P(r)$.

Using these background quantities, the coefficients $F(r)$ and $Q(r)$ from Eq. \ref{eq:H_equation} can be explicitly expressed as:
\begin{align}
    F(r) &= \left(1-\frac{2GM(r)}{r}\right)^{-1} \left[1 - 4\pi G r^2 (\epsilon(r) - P(r))\right] \\
    Q(r) &= \left(1-\frac{2GM(r)}{r}\right)^{-1} \left[ 4\pi G \left(5\epsilon(r) + 9P(r) + \frac{\epsilon(r) + P(r)}{v_s^2(r)}\right) - \frac{6}{r^2} - (\nu'(r))^2 \right]
\end{align}
where $v_s^2 = dP/d\epsilon$ is the local speed of sound squared, which directly reflects the stiffness of the equation of state. These expressions connect the microscopic properties of the EOS (via $\epsilon, P, v_s^2$) with the macroscopic gravitational structure of the star (via $M(r), \nu(r)$), and are crucial for calculating the tidal deformability.

\printbibliography 

\end{document}